\documentclass[fleqn,usenatbib]{mnras}

\usepackage{newtxtext,newtxmath}

\usepackage[T1]{fontenc}

\DeclareRobustCommand{\VAN}[3]{#2}
\let\VANthebibliography\thebibliography
\def\thebibliography{\DeclareRobustCommand{\VAN}[3]{##3}\VANthebibliography}

\usepackage{graphicx}	
\usepackage{amsmath}	
\usepackage[dvipsnames]{xcolor} 
\usepackage[flushleft]{threeparttable} 
\usepackage{float}
\usepackage{booktabs} 
\usepackage{physics} 
\usepackage{subcaption}
\usepackage[normalem]{ulem}
\usepackage{comment}

\title[Projected DF]{How much can we learn from resolved stellar kinematics of galactic haloes using action-based dynamical models?}

\author[P. Gherghinescu et al.]{
Paula Gherghinescu$^{1}$,\thanks{E-mail: p.gherghinescu@surrey.ac.uk}
Eugene Vasiliev$^{2}$,
Payel Das$^{2}$,
Justin Read$^{2}$ \\
$^{1}$Institute for Computational Cosmology, Department of Physics, Durham University, South Road, Durham, DH1 3LE, United Kingdom\\
$^{2}$University of Surrey, Guildford, GU2 7XH, United Kingdom
}

\date{Accepted XXX. Received YYY; in original form ZZZ}

\begin{document}
\label{firstpage}
\pagerange{\pageref{firstpage}--\pageref{lastpage}}
\maketitle

\begin{abstract}
\noindent Dynamical models are used to study dark matter (DM) in galaxies, how galaxies assemble through mergers, and to test galaxy formation models. Despite its widespread use, there has been no systematic study quantifying how much information can be obtained from just two on-sky positions and line-of-sight velocities, which are typically available for nearby external galaxies. In this work, we introduce axisymmetric, action-based dynamical models that use the positions and velocities of stellar halo stars to jointly constrain the total mass distribution of galaxies and the underlying DM component, as well as the stellar halo phase-space distribution. We rigorously test the method using both idealised equilibrium galaxy mocks and cosmological hydrodynamical simulations from the Auriga suite, systematically assessing how its performance degrades as the available phase-space information is progressively reduced. We further examine the impact of galaxy inclination, modelling assumptions, and methodological systematics on the recovered mass profiles. A crucial development in this work is the improved marginalisation of the model likelihood over missing phase-space dimensions. Our models successfully recover the total and DM mass distributions, as well as the kinematic properties of the stellar tracers, within the derived confidence intervals. However, we find that with limited (3D or 4D) phase-space information, the flattening of the DM halo cannot be constrained with any degree of certainty. Nevertheless, the recovered mass profile is insensitive to the flattening.  This finding is independently validated by Schwarzschild modelling tests.

\end{abstract}




\section{Introduction}

In the standard $\Lambda$CDM cosmological paradigm, galaxies are formed hierarchically and are dominated by dark matter (DM) at large scales. Stellar haloes and other tracers in the outer regions of galaxies, such as globular clusters, record the galaxy assembly history and are important probes of its gravitational potential. The mass profiles of DM haloes vary between different DM models \citep[see e.g.][and references therein]{Bechtol+2022}, but their shapes can also be used to constrain the physical nature of DM \citep{Dave+2001, Chua+2021}. The distribution of baryonic halo tracers in the position--velocity phase space provides a pathway for studying galaxy formation and properties of DM haloes through dynamical modelling.

\par Thanks to the Gaia satellite \citep{GaiaPaper2016}, we now have an accurate 6D phase-space map of our own Galaxy. However, our MW is only one of the many disc galaxies in the Universe. Furthermore, our efforts to study the entire halo are hindered by our position inside the Galaxy. This makes studying our neighbour galaxies the next step in bridging near-field and far-field galactic dynamics studies. For example, our closest neighbour, the Andromeda Galaxy (M31), is the most similar system to the Milky Way within the Local Group, both in mass and morphology. Its proximity ($\sim 780$ kpc) and edge-on orientation (inclination $i\sim 77^\circ$) offer a panoramic view and make it an ideal candidate for studying stellar haloes of external disc galaxies. Beyond M31, other Local Group members—including dwarf spheroidals and smaller satellites—offer insights into a broader range of galaxy types and mass scales, allowing us to place the MW in a wider cosmological context.

However, as we observe beyond the MW, the amount of astrometric information is limited. In contrast to the MW, where we can have up to full 6D phase-space information for individual stars (i.e. $x,y,z,v_{x},v_{y},v_{z}$), for Local Group galaxies, we are often limited to 3D phase-space information, i.e. position on the sky and line-of-sight (LOS) velocities for individual stars or tracers. This means that when studying external galaxies, we are inevitably dealing with reduced information regarding the system \citep[e.g.,][]{Battaglia&Nipoti2022}.

It is well known that attempts to determine the mass distribution of a spherical system from just the second moment of LOS velocities (such as provided by Jeans equations) are thwarted by the mass--anisotropy degeneracy \citep{Binney&Mamon1982}, which can be alleviated by the use of higher-order moments \citep{Merrifield&Kent1990,Richardson&Fairbairn2014,Read&Steger2017} or the full LOS velocity profiles \citep{Dejonghe&Merritt1992}, and of course by the addition of proper motions (PM; \citealt{Strigari+2007,BanaresHernandez+2025}). When the spherical assumption is relaxed, a new degeneracy with flattening/inclination is possible. It is not known to what extent the use of full 3D phase-space distributions can ameliorate the degeneracy in axisymmetric, dispersion-dominated systems (see e.g.\ Chapter 3.4 in \citealt{Merritt2013} for a review).

\par In this work, we investigate the effect of missing phase-space information when dynamically modelling external galaxies, for which resolved stellar populations are available. We build upon the work from \citealt{Gherghinescu+2024}, where we introduced action-based equilibrium dynamical models for the stellar halos of M31-like galaxies within the Auriga simulation suite \citep{Grand+2017_AurigaSims}. We showed that our Bayesian model-fitting algorithm successfully recovered the total mass and DM distribution of the Auriga galaxies and the distribution function (DF) of the stellar halo, when considering full 6D phase-space knowledge of the kinematic tracers (i.e. stellar halo stars). In this paper, we extend the previous study to include the effects introduced by having realistic missing data information, i.e. we only have access to position on the sky, but not the distance, and one or more velocity components: LOS velocity (i.e., 3D phase-space information, our main and most realistic scenario), two PM (4D), or both $v_\mathrm{LOS}$ and PM (5D). Furthermore, we investigate the impact of the galaxy inclination on the recovered properties.

\par This paper is organised as follows. In Section~\ref{sec:method}, we introduce our dynamical model and the Bayesian fitting routine used to constrain the model parameters and therefore recover the galaxy total and DM mass distribution, as well as the phase-space distribution of the stellar tracers. In Section~\ref{sec:mock_tests}, we introduce an idealised equilibrium mock galaxy and investigate the effects of missing coordinates, as well as the inclination angle, using the modelling framework presented earlier. We focus on the 3D phase space information scenario, but put them into context by comparing the results for 4D, 5D, and 6D phase space availability. Next, in Section~\ref{sec:auriga_tests}, we do the same for a realistic galaxy from the Auriga simulation suite (Auriga 23). In Section~\ref{sec:schwarzschild} we verify our results for idealised mocks using a complementary approach (Schwarzschild orbit-superposition modelling). Finally, we summarise and discuss our findings in Section~\ref{sec:discussion_conclusions}.

\section{Method} \label{sec:method}
In this Section, we present the action-based dynamical modelling approach, which is based on \cite{Read+2021} and \cite{Gherghinescu+2024}, and adapted for the case of incomplete data, i.e. position on the sky and LOS information only (although we also explore the case of having only position and PM, or position and all three velocity components).

\par Our dynamical models include a multicomponent galactic potential (bulge + disc + DM halo) and an action-based double power law DF for the stellar halo. The model is implemented using the \texttt{AGAMA} package \citep{Vasiliev2019_Agama}.  \par Actions $({J_{r},J_{\phi},J_{z}})$ are canonical coordinates which are also integrals of motion and characterize stellar orbits. They also have a straightforward interpretation: ${J_{r}}$ is a measure of the orbit eccentricity, ${J_{z}}$ describes the wandering above and below the galactic symmetry plane, and ${J_{\phi}}$ quantifies the degree of prograde or retrograde motion.

\subsection{The gravitational potential} \label{Method:potential}
The gravitational potential is assumed to be an oblate axisymmetric composite potential comprising a bulge, disc, and DM halo. Density profiles are defined for each of these components from which the potential is recovered through the Poisson equation. The density of the disc follows an exponential disc profile:
\begin{equation}
    \rho_\mathrm{disc}(r) = \Sigma_{0}\exp\bigg( -\frac{R}{R_\mathrm{disc}}  \bigg) \times \frac{1}{4|h_\mathrm{disc}|} \mathrm{sech^{2}}\bigg(\bigg |\frac{z}{2h_\mathrm{disc}} \bigg|\bigg),
\end{equation}
with $\Sigma_{0}$ being the surface density at $R=0$, $R_\mathrm{disc}$ the disc scale radius, and $h_\mathrm{disc}$ its scale height.

The density of the bulge follows a spherical S\'ersic profile, which is the deprojection of
\begin{equation}
    \Sigma(R) = \Sigma_{0}\exp[-b_{n}\biggl(\frac{R}{R_\mathrm{bulge}} \biggr)^\frac{1}{n_\mathrm{bulge}}],
\end{equation}
where $R = \sqrt{x^2 + y^2}$ with $(x,y)$ the Cartesian coordinates of the star in the sky plane, $\Sigma_{0}$ is the surface density normalisation, $R_\mathrm{bulge}$ is the scale radius, $n_\mathrm{bulge}$ is the Sérsic index, and $b_{n}\approx 2n_\mathrm{bulge}-1/3$ is the root of $\Gamma(2n)=2\gamma(2n, b_n)$. 
\par The DM density follows a flattened NFW profile \citep{Navarro+1997_NFW}:
\begin{equation}
    \rho_{\text{DM}}(R,z) = \frac{\rho_{0,\text{DM}}}{(\tilde r/r_\mathrm{DM})\;\left(1+\tilde r/r_\mathrm{DM}\right)^{2}},
\end{equation}
where $\rho_{0,\text{DM}}$ is the normalisation density and $r_\mathrm{DM}$ is the scale radius. The elliptical radius $\tilde r = \sqrt{x^2+y^2+(z/q_\mathrm{DM})^2} = \sqrt{R^2+(z/q_\mathrm{DM})^2}$ takes into account the flattening of the halo through the parameter $q_\mathrm{DM}$ (axis ratio $z/R$). The total potential of the galaxy is the sum of the potentials generated by the individual components,
\begin{equation}
    \Phi(R,z) = \Phi_{\mathrm{disc}}(R,z) + \Phi_{\mathrm{bulge}}(R,z) + \Phi_{\mathrm{DM}}(R,z),
\end{equation}
where $\Phi_{\mathrm{disc}}(R,z), \Phi_{\mathrm{bulge}}(R,z),$ and $\Phi_{\mathrm{DM}}(R,z)$ are the potentials of the disc, bulge, and DM halo, respectively.

\subsection{The stellar halo DF} \label{Method:dfs}
To model the action-space distribution of stars in the stellar halo, we use a generalisation of the DF proposed by \citet{Posti+2015_sphDF} for spheroidal galactic components. This generalised DF has been implemented in \texttt{AGAMA} and has the form:

\begin{align} 
\begin{split}
   & f(\vb*{J})  = \frac{M_{0}}{(2\pi J_{0})^{3}}\;
   \biggl[ 1 + \left( \frac{J_{0}}{h(\vb*{J})} \right)^\eta \biggr]^{\alpha/\eta}\;
   \biggl[ 1 + \left( \frac{g(\vb*{J})}{J_{0}} \right)^\eta \biggr]^{-\beta/\eta} \\
   &\mbox{}\qquad\times  \biggl(1+\chi \tanh\frac{J_{\phi}}{J_{\phi,0}}\biggr), \\ 
& \text{where}\,\,g(\vb*{J}) = g_{r}J_{r} + g_{z}J_{z} + (3- g_{r}-g_{z})\left | J_{\phi} \right |, \\
& \text{and}\,\;\;\;\,h(\vb*{J}) = h_{r}J_{r} + h_{z}J_{z} + (3- h_{r}-h_{z})\left | J_{\phi} \right |.
\end{split}
\label{eq:double_power_law_df}
\end{align}
Here $M_{0}$ is a normalisation parameter with units of mass (which differs from the total mass of the component); $\alpha$ and $\beta$ are the power-law indices controlling the behaviour of the density profile at small and large radii, respectively; $J_{0}$ is the scale action of the transition between the inner and outer regimes; and $\eta$ is the steepness of this transition. The functions $h$ and $g$ are linear combinations of actions specified by the mixing coefficients $\{g,h\}_{r,z,\phi}$, whose sum is fixed to 3 (so that there are only two independent coefficients). They have the overall effect of controlling the flattening and anisotropy of the density and velocity ellipsoids in the inner and outer regions, respectively. Finally, $\chi$ and $J_{\phi,0}$ control the model's rotation.

\subsection{Bayesian fitting of the stellar halo model}
In our fitting procedure, the structural parameters of the baryonic components (disc and bulge) are kept fixed, but their overall mass normalisation is a free parameter, and for the DM component, all parameters are free. We define $\Upsilon\equiv M/L$ as the mass-to-light ratio, which converts the observed luminosity of the disc and bulge into stellar mass (i.e., the overall multiplicative factor for the bulge and disc masses). We assume that we can determine the structural parameters (scale radii, height, S\'ersic index, etc.) of baryonic components from photometry, which is a reasonable assumption, especially for external galaxies in our Local Group. The parameters defining the model are $\vb*{P}=(\alpha, \beta, \eta, J_{0},h_{r}, g_{r}, h_{z}, g_{z}, \chi, J_{\phi,0}, \Upsilon, \rho_{0,\rm DM}, R_\mathrm{DM}, q_\mathrm{DM})$. For the tests described in Section~\ref{sec:mock_tests}, we also have the inclination $i$ as a free parameter.

We fit the model using a Bayesian approach, where we define the individual likelihood $\ell_{i}$ of each star $i$ at its observed phase-space coordinates $\boldsymbol w_i \equiv \{\boldsymbol x_i, \boldsymbol v_i\}$ as the value of the DF at that point, normalized by the total mass $\mathcal{N}$ of the stellar halo component in the specified model:
\begin{equation}  \label{eq:likelihood_star}
    \ell_{i}(\vb*{J}) = \frac{f\big(\boldsymbol J(\boldsymbol w_i ;\;\Phi)\big)}{\mathcal{N}}, \qquad \mathcal{N} \equiv \iiint f(\boldsymbol J)\,d^3J. 
\end{equation}
Thus the total likelihood of the model, $\mathcal{L} = P(\mathcal{D}|\vb*{P})$, is:
\begin{equation}  \label{eq:likelihood_model}
    \mathcal{L} = \prod_{i}^{n_{*}} \ell_{i} \;\Rightarrow\; \log\mathcal{L} = \sum_{i}^{n_{*}}\log\ell_{i} = \sum_{i}^{n_{*}}\log\frac{f(\boldsymbol{J}_{i})}{\mathcal{N}},
\end{equation}
where the sum is over all the $n_{*}$ stars in the dataset.

According to Bayes' law, the posterior probability is given by:
\begin{equation}
    P(\vb*{P}|\mathcal{D}) = \frac{P(\mathcal{D}|\vb*{P})\; P(\vb*{P})}{P(\mathcal{D})},
\end{equation}
where $P(\mathcal{D}|\vb*{P})$ is the likelihood (i.e., $\mathcal{L}$ from equation \ref{eq:likelihood_model}), $P(\vb*{P})$ is the prior, and $P(\mathcal{D})$ is the evidence.
The prior conditions (using flat priors on the listed parameters) are:
\begin{itemize}
    \item $0 < \alpha < 3$
    \item $3 < \beta < 12$
    \item $1/3 < \eta < 3$
    \item $h_{r}>0$, $h_{z}>0$ and $3-h_{r}-h_{z}>0$
    \item $g_{r}>0$, $g_{z}>0$ and $3-g_{r}-g_{z}>0$
    \item $-1 \leq \chi \leq 1$
    \item $3 \leq \log_{10} \big( {J_{0}} / [\mathrm{kpc\,km\,s}^{-1}] \big) \leq 6$
    \item $1 \leq \log_{10} \big( {J_{\phi,0}} / [\mathrm{kpc\,km\,s}^{-1}] \big) \leq 8$
    \item $6 \leq \log_{10} \big( \rho_{0,\rm DM} / [\mathrm{M_{\odot}kpc^{-3}}] \big) \le 9$
    \item $0 \leq \log_{10} \big( r_\mathrm{DM} / \mathrm{kpc} \big) \leq 2$
    \item $0.2 \leq q \leq 1$
\end{itemize}

\noindent The evidence integral is given by:
\begin{equation}
    P(\mathcal{D}) = \int P(\mathcal{D},\vb*{P}) \mathrm{d}\vb*{P}. 
\end{equation}
Given a survey, it would return the probability of all stars being part of that survey. In our case, since all data points are part of the dataset, we can leave out the integral, as it will only shift up or down the log-posterior distribution by the same value.

\par The log-posterior distribution is explored using Markov Chain Monte Carlo (MCMC) sampling via the \texttt{emcee} Python package \citep{Foreman-Mackey+2013_MCMCHammer}.

\subsection{Treatment of measurement uncertainties and missing dimensions}  \label{sec:marginalisation}
\par In the case of missing or incomplete data, we need to marginalise the likelihood (\ref{eq:likelihood_star}) over the missing coordinates. Likewise, in the case of measurement uncertainties in the data, the likelihood has to be convolved with the error distribution. However, in this study, we work with mock tests and simulations in which we assume no errors.
We denote the coordinates in the intrinsic reference frame of the model as $xyz$ (with $z$ being the axis perpendicular to the disc, which is also the minor axis of the DM halo), and the coordinates in the observer's reference frame as $XYZ$; the latter is still centred on the galaxy, but rotated by an angle $i$ about the $x\equiv X$ axis, such that $Z$ is the direction along the line of sight and $X,Y$ are the coordinates in the sky plane.
In the case of missing distance $Z$ and PM $v_{X,Y}$, the marginalised log-likelihood becomes:
\begin{equation}  \label{eq:marginalisation}
    \log \mathcal{L} = \sum_{i=1}^{n_{\star}} \log \tilde{\ell_i}, \quad
    \tilde{\ell_i} \equiv \int_{-\infty}^{+\infty}\mathrm{d}Z \int_{-\infty}^{+\infty}\mathrm{d}v_{X} \int_{-\infty}^{+\infty}\mathrm{d}v_{Y} \frac{f(\boldsymbol{J}_{i})}{\mathcal{N}} .
\end{equation}
Similar expressions can be written for other combinations of missing coordinates or velocities.

The marginalisation integral (\ref{eq:marginalisation}) can be computed numerically using the multidimensional adaptive integration method implemented in the \texttt{cubature} library \citep{Johnson2005_Cubature}. This is the method used by the \texttt{projectedDF} routine in \texttt{Agama}; it is usually very accurate but also too expensive to use in MCMC runs. Alternatively, this integral can be estimated using a Monte Carlo approach; that is, summing over $n_{\mathrm{MC}}$ points sampling the missing coordinates: 
\begin{equation}
  \tilde{\ell_i} \approx \frac{1}{n_{\mathrm{MC}}} \sum_{k=1}^{n_\mathrm{MC}} \frac{f\big(\boldsymbol J(\boldsymbol w_{i,k})\big)}{\mathcal N}.
\end{equation}
A few additional steps are needed to make it work in practice:
\begin{enumerate}
\item As argued by \cite{McMillan+2013}, in order to eliminate or at least substantially reduce the Poisson noise associated with Monte Carlo sampling, one should keep the set of samples fixed throughout the entire MCMC process. In this way, even if the Monte Carlo integral deviates from the true value, the resulting error in the likelihood is similar for all models in the MCMC set, so does not significantly affect their relative odds.
\item The missing coordinates can span a range that is either infinite ($Z$) or at least large ($|v_{X,Y}|\le v_{\mathrm{escape}} \gg \sigma$), so sampling them uniformly is impractical. \citet{Read+2021} introduced an importance sampling procedure, where $Z$ is drawn from a conditional 1d distribution $\rho(Z|R)$ with the 3d density profile $\rho(r)$ obtained by deprojecting the surface density of the observed sample, and $v_{X,Y}$ are drawn from a bell-shaped 2d Laplacian distribution with a width estimated from the dispersion of LOS velocities $\sigma_Z$. To compensate for the non-uniform distribution of these draws, the weight of each sample in the Monte Carlo integral is inversely proportional to the sampling distribution function. Ideally, if these distributions closely match the actual values of the DF $f$ at these points, this minimises the error in the Monte Carlo estimate; in practice, since the DF is unknown \textit{a priori}, we may only partially achieve this objective.
\end{enumerate}
In the present study, we improved the marginalisation and importance sampling procedure in several aspects:
\begin{enumerate}  \setcounter{enumi}{2}
\item Originally, the deprojection was performed by fitting a smoothing spline $\Sigma(R)$ to the distribution of stars in projected radius (see Appendix~A2.5 in \citealt{Vasiliev2018_Agamaext}), then the spherical density profile $\rho(r)$ was constructed using the Abel inversion, and finally the $Z$ coordinate was sampled from the probability distribution $p(Z\;|\;R) \equiv \rho(\sqrt{R^2+Z^2}) \big/ \Sigma(R)$. The last step was rather inefficient, so in the current implementation, after obtaining the spline for $\Sigma(R)$, we approximate it by a sum of $\mathcal O(10)$ Gaussians, for which the deprojection and sampling $Z$ from the conditional distribution can be performed analytically. Interestingly, direct fitting of the original distribution of stars in $R$ by a sum of Gaussians results in a noisier solution than an intermediate smoothing spline followed by a multi-Gaussian approximation.
\item The actual DF in the model may result in a velocity distribution that is strongly peaked at small $|v|$, or in other cases, has a broader tails than a Gaussian function. In either case, a simple Gaussian may not be a good choice for the sampling function of missing velocity components, and after some experiments, we chose the following functional forms for the sampling distribution, which are both peaked and fat-tailed. For the two sky-plane velocities, we sample their total magnitude from $p\big(|v|\big) \propto (v^2+\varsigma^2)^{-3/2}$ and then split it between $v_X,v_Y$ isotropically. A missing $v_Z$ is sampled from $p(v_Z) \propto |v_Z|^{-1/2}\,(v_Z^2+\varsigma^2)^{-5/4}$. In both cases, the width $\varsigma$ of these sampling distributions is taken to be twice the 1d velocity dispersion of input points (this value was chosen after some experimentation to maximise the efficiency for near-Gaussian distributions). With these choices, a small fraction ($\lesssim 10\%$) of samples usually exceeds the escape velocity of the best-fit model (thus the DF for them is zero), but this minor inefficiency is tolerable and avoids more serious biases that may occur in case of sampling from a Gaussian that turns out to be a poor fit to the actual velocity distribution.
\item The most dramatic improvement comes from replacing pseudo-random numbers in the sampling procedure with quasi-random (low-discrepancy) ones; specifically, the scrambled Halton sequence \citep{Owen2017_QRNG}. It is well known that for well-behaved integrands, this results in the error of the Monte Carlo estimate decreasing as $\sim\!N_\mathrm{MC}^{-1}$ instead of $\sim\!N_\mathrm{MC}^{-1/2}$ as in the case of (pseudo)-random sampling. We have empirically verified that for $n_\mathrm{MC} \gtrsim 10^3$, the difference in the total $\log\mathcal L$ for all stars between the Monte Carlo integration and \texttt{projectedDF} is usually $\lesssim 1$, sufficiently small not to cause any bias in the parameter inference. 
\end{enumerate}

\section{Application to mock galaxies with different inclinations} \label{sec:mock_tests}

A first consistency test of the DF fitting method is its performance in cases when the mock data were generated from an equilibrium model with known parameters, and the range of possible models in the fit includes the true one. The equilibrium model is constructed as described in Section \ref{sec:method}, and the mock galaxy's density components and stellar halo parameters are specified in Table \ref{tab:mock_param}. The stellar halo component specified by the double-power-law DF does not contribute to the potential.

We create data sets of $n_{*}=1000$ stars from the DF of the galaxy model described above. This choice reflects the typical number of available tracers with LOS information in M31 (e.g., planetary nebulae, globular clusters) as well as in dSph, while also keeping the computational cost of the likelihood evaluation manageable.

To quantify stochastic effects, we generate four independent realisations of these data sets; specifically, we draw a much larger initial number of samples $\mathcal O(10^6)$ and then retain several non-overlapping subsets of $n_{*}$ points.
Although all sets originate from the same parent DF and are therefore expected to share the same statistical properties in the limit of large samples, finite sampling introduces Poisson noise. These multiple realisations thus provide a straightforward way to assess the sensitivity of our modelling procedure to such systematics.

\begin{figure}
    \includegraphics[scale=0.25]{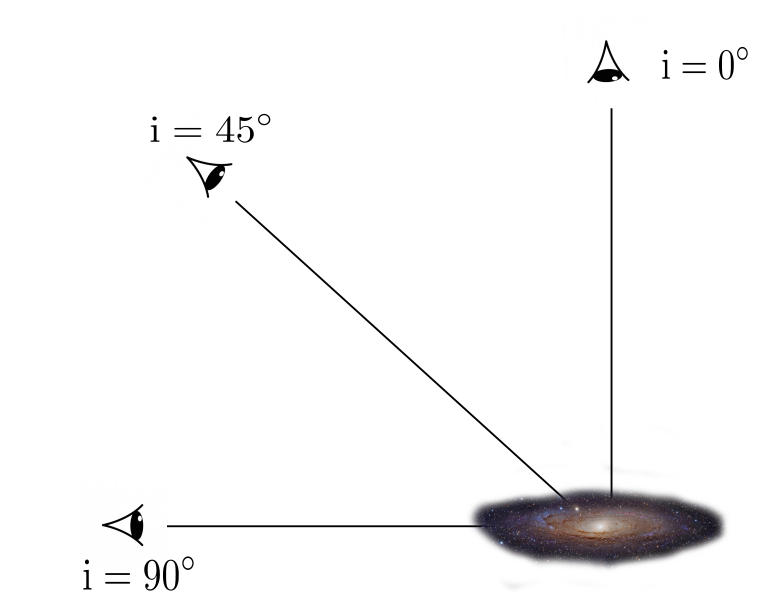}
    \caption{The different orientations investigated for the galaxy: $i=0^{\circ}$ (face-on), $i=45^{\circ}$, and $i=90^{\circ}$ (edge-on) with respect to the axisymmetry plane which is defined by the orientation of the galactic disc.}
    \label{fig:inclinations}
\end{figure}

Finally, we examine how the inclination affects the parameter recovery. For each of the four base data sets described above, we produce  versions at three different inclinations (see Figure \ref{fig:inclinations}):  ${i=0^{\circ}}$ (disc is face-on), ${i=45^{\circ}}$ (intermediate), and ${i=90^{\circ}}$ (disc is edge-on).

\begin{table}
  \caption{\label{tab:mock_param} The parameters values of the mock galaxy.}
  \centering 
  \begin{threeparttable}
    \begin{tabular*}{0.8\linewidth}{@{\extracolsep{\fill}}ccc}
\midrule
    Component  & Parameter & Value mock \\
     \midrule
DM halo potential  &   
$\begin{aligned}
    \rho_{0,\mathrm{DM}}\\
    r_\mathrm{DM} \\
    q_\mathrm{DM}
\end{aligned}
$        &  
$\begin{aligned}
    & 1.1\times 10^{7} \: M_{\odot}\,\mathrm{kpc^{-3}} \\
    & 17 \textrm{ kpc} \\
    & 0.7
\end{aligned}$

\\
    \cmidrule(l  r ){1-3}
    Bulge potential & $ \begin{aligned}
    M_{\mathrm{bulge}} \\
    R_{\mathrm{bulge}} \\
    n_{\mathrm{bulge}} \\
    \end{aligned} $ &
    $\begin{aligned} 
    & 0.93 \times 10^{10} \: M_{\odot} \\
    & 0.877 \textrm{ kpc}\\
    & 2.0 \\
    \end{aligned}$

\\
    \cmidrule(l  r ){1-3}
    Disc potential & $ \begin{aligned}
    M_{\mathrm{disc}} \\
    R_{\mathrm{disc}} \\
    h_{\mathrm{disc}}
    \end{aligned} $ &
    $\begin{aligned} 
    & 6.648 \times 10^{10} \: M_{\odot} \\
    & 2.591 \textrm{ kpc}\\
    & 0.518 \textrm{ kpc}
    \end{aligned}$

\\
    \cmidrule(l  r ){1-3}
     Stellar halo DF & $ \begin{aligned}
    \alpha \\
    \beta \\
    \eta \\
    J_{0} \\
    h_{r} \\
    h_{z} \\
    g_{r} \\
    g_{z} \\
    \chi \\
    J_{\phi,0}
    \end{aligned} $ &
    $\begin{aligned} 
    & 2.5 \\
    & 5.5 \\
    & 1.0 \\
    & 8000 \: \mathrm{kpc\,km\,s^{-1}} \\
    & 0.75 \\
    & 1.7 \\
    & 0.88 \\
    & 1.1 \\
    & 0.5 \\
    & 1000 \: \mathrm{kpc\,km\,s^{-1}} \\
    
    \end{aligned}$  

\\ 
    \midrule
    \end{tabular*}
\end{threeparttable}

\end{table}

\subsection{Mass distribution} \label{subsec:mass_distribution_idealised_mocks}
\begin{figure}
    \centering
    \includegraphics[width=0.95\linewidth]{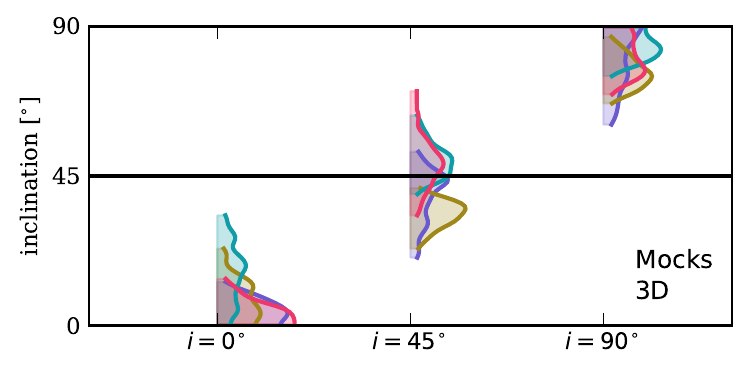}
    \caption{The recovery of the inclination angle for the idealised mock tests with 3D phase space availability. Different coloured lines correspond to fits to four independent data realisations.}
    \label{fig:inclination_mock_test_3d}
\end{figure}

\begin{figure*}
    \centering
    \includegraphics[width=1\textwidth]{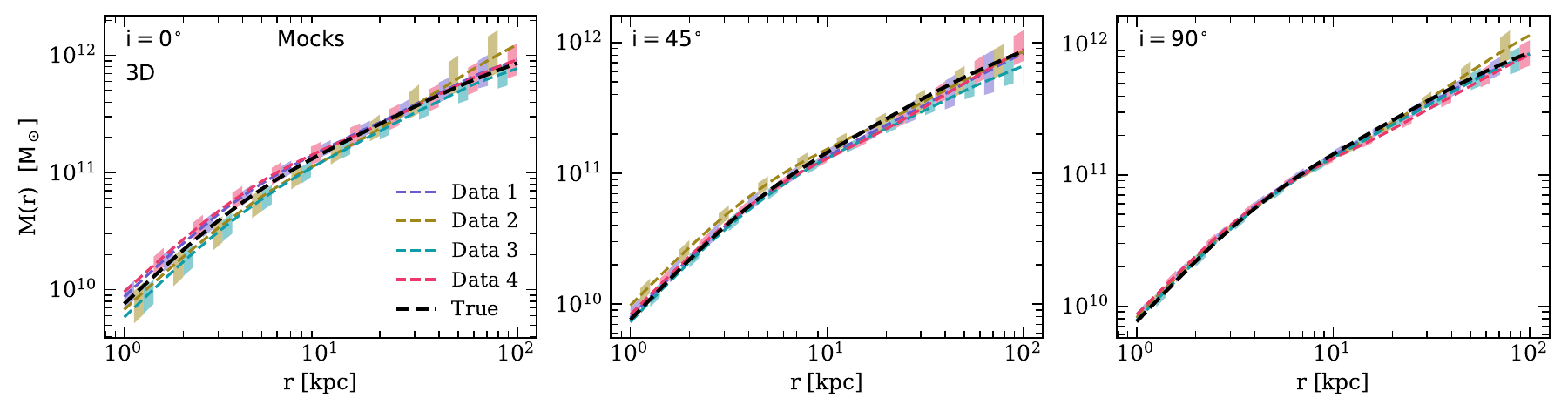}
    \caption{Total enclosed mass profiles of the best-fit models (coloured lines) compared to the true mass distribution of the mock galaxy (black dotted line) for inclination angles (a) ${i=0^{\circ}}$, (b) ${i=45^{\circ}}$, and (c) ${i=90^{\circ}}$, where the tracers used to constrain the potential have 3D phase-space information. The different coloured lines correspond to fits to four independent data realisations. Shaded regions indicate the 1$\sigma$ uncertainties.}
    \label{fig:encmass_total_mock}
\end{figure*}

Figure \ref{fig:encmass_total_mock} shows the total enclosed mass profiles of the recovered models compared to the true mass distribution (black dotted line) for the four data realizations. Overall, the total mass is recovered well, with the true profile consistently lying within the confidence intervals across all inclination scenarios. We nevertheless observe a systematic dependence on inclination: the recovery is least precise in the face-on case ($i=0^{\circ}$) and improves progressively towards the edge-on configuration ($i=90^{\circ}$), particularly in the inner regions of the galaxy. The inclination itself is recovered quite accurately (within 10--15$^\circ$) for all datasets (see Fig. \ref{fig:inclination_mock_test_3d}).

\begin{figure*}
    \centering
    \includegraphics[width=1\textwidth]{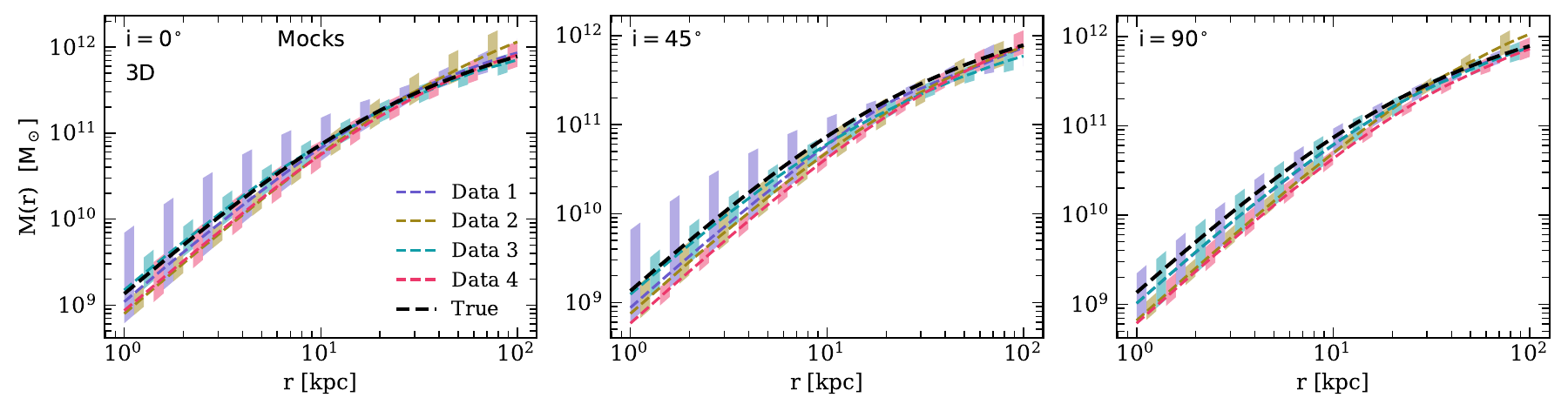}
    
    \caption{DM enclosed mass profiles of the best-fit models (coloured lines) compared to the true mass distribution of the mock galaxy (black dotted line) for inclination angles (a) ${i=0^{\circ}}$, (b) ${i=45^{\circ}}$, and (c) ${i=90^{\circ}}$, where the tracers used to constrain the potential have 3D phase-space information. The different coloured lines correspond to fits to four independent data realisations. Shaded regions indicate the 1$\sigma$ uncertainties.}
    \label{fig:encmass_dm_mock}
\end{figure*}

The recovery of the DM enclosed mass profiles also exhibits a systematic dependence on inclination (see Figure \ref{fig:encmass_dm_mock}). The largest uncertainties occur in the inner regions of the DM distribution, while the fits improve at larger radii. Notably, one particular realization (data set 1, shown in purple) displays especially large scatter in the inner profile, highlighting the impact of Poisson noise as an additional source of systematic uncertainty.

\par
\begin{figure}
\includegraphics[width=1\linewidth]{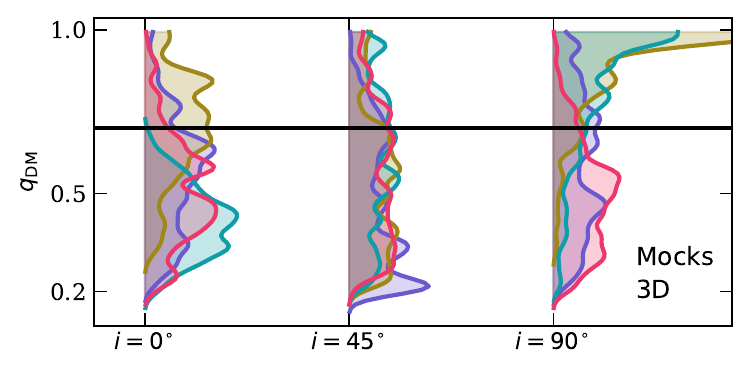}
  \caption{The posterior distribution of the DM flattening parameter $q_\mathrm{DM}$ for all inclinations and data set realisations (coloured lines): ${i=0^{\circ}}$, ${i=45^{\circ}}$, and ${i=90^{\circ}}$, where the tracers used to constrain the potential have 3D phase-space information. The different coloured lines correspond to fits to four independent data realisations. The solid black line is the true value of the parameter used to generate the mock galaxies ($q_\mathrm{DM,\, true}=0.7$). } 
  \label{fig:q_dm_3d}
\end{figure}

 While most parameters in the MCMC fits are generally well recovered, the axis ratio $q_{\mathrm{DM}}$ of the DM halo is the most challenging to constrain in the fully 3D case. As shown in Figure \ref{fig:q_dm_3d}, the posterior distribution of the recovered $q_{\mathrm{DM}}$ is relatively flat, indicating that the fits are not strongly sensitive to the choice of DM flattening. This behaviour is already apparent in our 4D experiments discussed in Section~\ref{sub:4d5d6d_mock} (see Figure \ref{fig:q_dm_4D_5D_6D}). In particular, the axis ratio $q_{\mathrm{DM}}$ of the DM halo may be biased towards smaller values than the true value $q_{\mathrm{DM}}=0.7$, pushing against the lower bound of the allowed range ($q_{\mathrm{DM}}\ge 0.2$), as can be seen in the 3D case for the $i=45^{\circ}$ inclination for data set 1 (Figure \ref{fig:q_dm_3d}), as well as for data sets 2 and 4 in the 4D case for the $i=0^{\circ}$ inclination (see Figure \ref{fig:q_dm_4D_5D_6D}). However, in the higher-dimensional cases (5D and 6D), the posterior increasingly concentrates around the true value of $q_{\mathrm{DM}} = 0.7$.

\subsection{Kinematics of the tracers} \label{subsec:kinematics_idealised_mocks}

In the previous subsection \ref{subsec:mass_distribution_idealised_mocks}, we discussed the ability of our dynamical model to recover the mass distribution of the galaxy using stellar halo tracers. In this subsection, we look into the recovery of the kinematic properties of the tracers used to constrain the galactic potential for all four generated data sets. Specifically, we analyse the first and second moments of the velocity distributions in each coordinate, which are computed from the DF using the \texttt{GalaxyModel.moments} routine. The non-trivial moments are the mean rotation velocity $\overline{v_\phi}$ and the three second moments $\overline {v_{R,z,\phi}^2}$. 

\par
\begin{figure*}
  \centering
  \includegraphics[width=\linewidth]{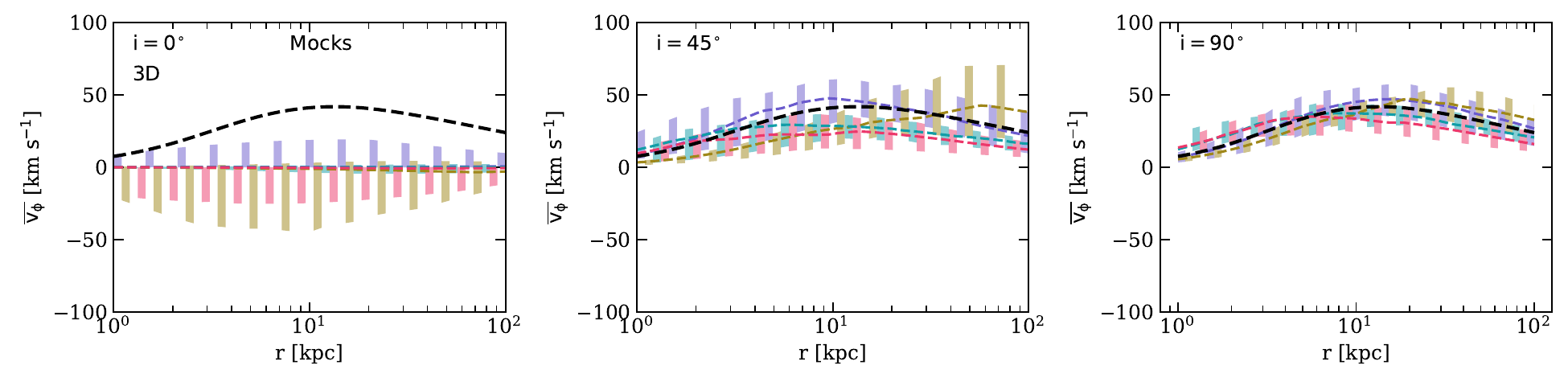}
  \caption{The true vs recovered median ${v_{\phi}}$ velocity component (streaming motion) for all four data sets for all the inclination experiments (a) ${i=0^{\circ}}$, (b) ${i=45^{\circ}}$, and (c) ${i=90^{\circ}}$ for the mock galaxy test, where the tracers used to constrain the potential have 3D phase-space information. The different coloured lines correspond to fits to four independent data realisations. This quantifies the overall degree of rotation about the axisymmetry axis of the stellar halo at different radii. It can be seen that it is best recovered for edge-on orientation and no rotation signal is detected for the face on orientation, as expected.}
  \label{fig:med_vel_vphi_mock}
\end{figure*}

 Figure \ref{fig:med_vel_vphi_mock} shows the recovery of the $\overline{v_{\phi}}$ streaming motion, which traces the overall rotation of the stellar halo component. As expected, the rotation signal is most accurately recovered in the edge-on case, where the $v_{\phi}$ component is aligned with the LOS. In contrast, no rotation signal is recovered in the face-on case. The $i=45^{\circ}$ configuration is an intermediate scenario.

\par
\begin{figure*}
  \centering
    \begin{subfigure}{0.95\linewidth}
    \centering
      \includegraphics[width=\linewidth]{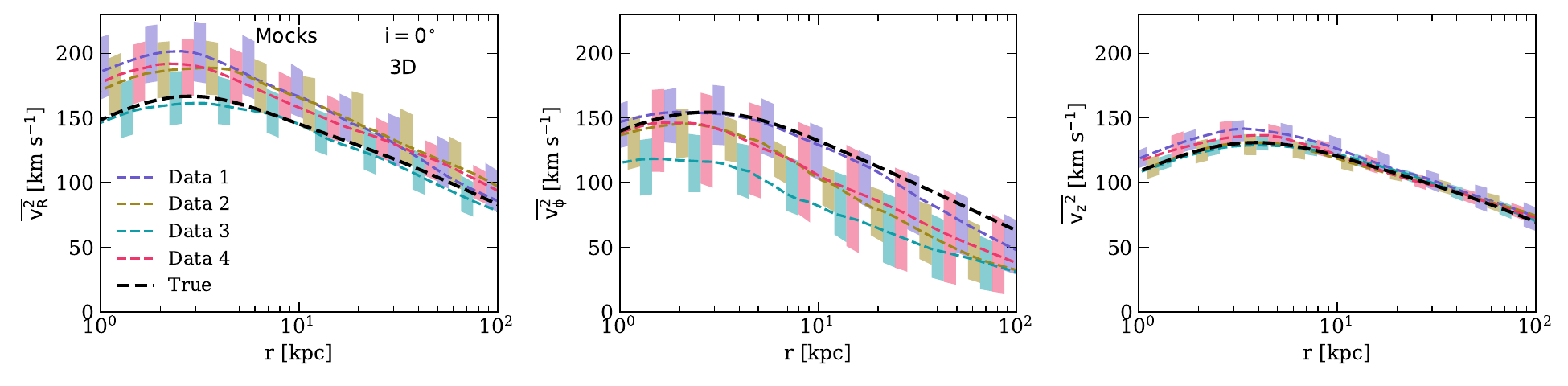}
      \caption{}
      \label{subfig:veldisp_cyl_i0_mock}
    \end{subfigure}\\
    \begin{subfigure}{0.95\linewidth}
        \centering
      \includegraphics[width=\linewidth]{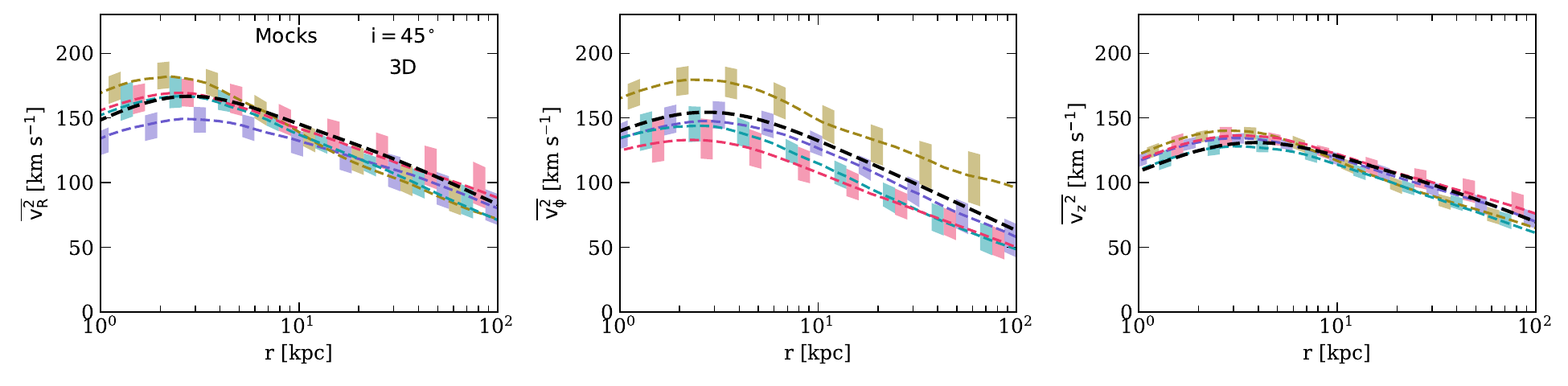}
      \caption{}
      \label{subfig:veldisp_cyl_45_mock}
    \end{subfigure}\\
    \begin{subfigure}{0.95\linewidth}
        \centering
      \includegraphics[width=\linewidth]{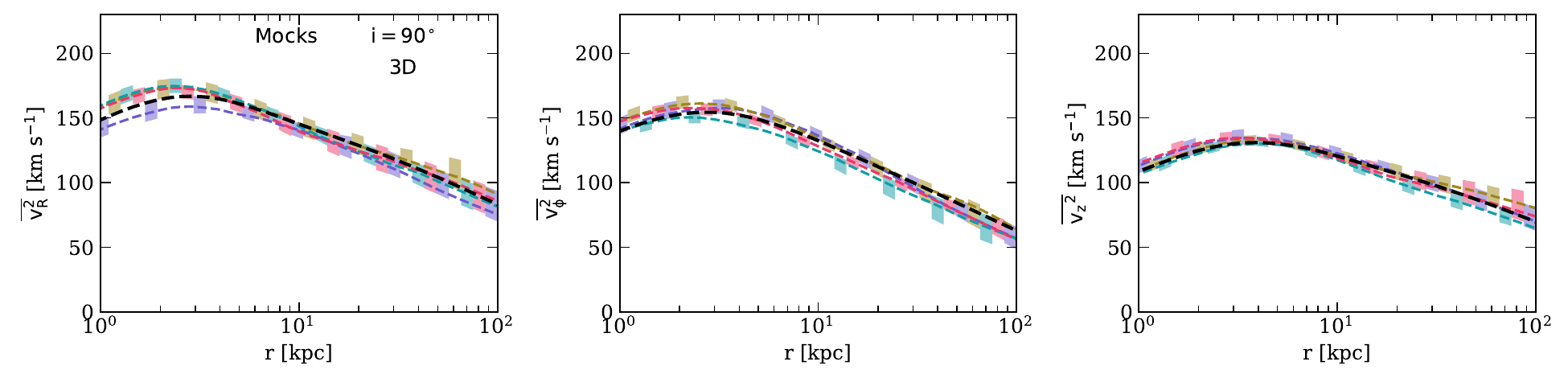}
      \caption{}
      \label{subfig:veldisp_cyl_i90_mock}
    \end{subfigure}    
  \caption{Velocity dispersions in cylindrical coordinates of the recovered models (coloured lines) compared to the true profile (black, dotted line) of the mock galaxy (black dotted line) for inclination angles (a) ${i=0^{\circ}}$, (b) ${i=45^{\circ}}$, and (c) ${i=90^{\circ}}$, where the tracers used to constrain the potential have 3D phase-space information. The different coloured lines correspond to fits to four independent data realisations. Shaded region indicate the 1$\sigma$ uncertainties.}
  \label{fig:veldisp_cyl_mock}
\end{figure*}

 Figure \ref{fig:veldisp_cyl_mock} compares the true velocity dispersions of the tracers (black lines) with the recovered profiles (coloured lines) in cylindrical coordinates. 
In the face-on case ($i=0^{\circ}$), the in-plane velocity components ($\overline{v^{2}_{R}}$ and $\overline{v^{2}_{\phi}}$) are not well recovered and exhibit large uncertainties. This is expected, since these components are oriented in the sky plane and do not contribute to the observed LOS velocity. The latter is aligned with the $z$ axis of the galaxy's intrinsic coordinate system, and therefore the corresponding velocity dispersions are recovered accurately. Conversely, in the edge-on case, the LOS velocity probes both in-plane velocity components $v^2_{R,\phi}$, so they are much better recovered.

\section{Application to Auriga 23 halo} \label{sec:auriga_tests}
We now apply our action-based dynamical model to a more realistic system — the Auriga 23 galaxy from the Auriga cosmological simulation suite \citep{Grand+2017_AurigaSims}. As described in Sections \ref{sec:method} and \ref{sec:mock_tests}, we use stellar halo stars as tracers of the gravitational potential. In this work, the stellar halo is defined as the accreted stellar component in the simulation. 

\par Four data sets are again constructed by randomly sampling $n_{*} = 1000$ stellar halo stars, ensuring uniform spatial coverage, such that we avoid biasing our data sample with stars contained only in substructures or clustered regions within the halo. An action-based dynamical model was previously applied to the stellar halo of Auriga 23 to constrain the total and DM mass distribution, as well as the stellar halo DF, using the full 6D phase-space information of the tracers \citep{Gherghinescu+2024}. This gives the 'true' underlying galaxy model of the simulated galaxy, which allows a direct assessment of the model’s recovery performance in the case of degraded 3D data. For the tests in this Section, we keep the inclination fixed to its true value.

\subsection{Mass distribution}

\begin{figure*}
    \centering
    \includegraphics[width=1\textwidth]{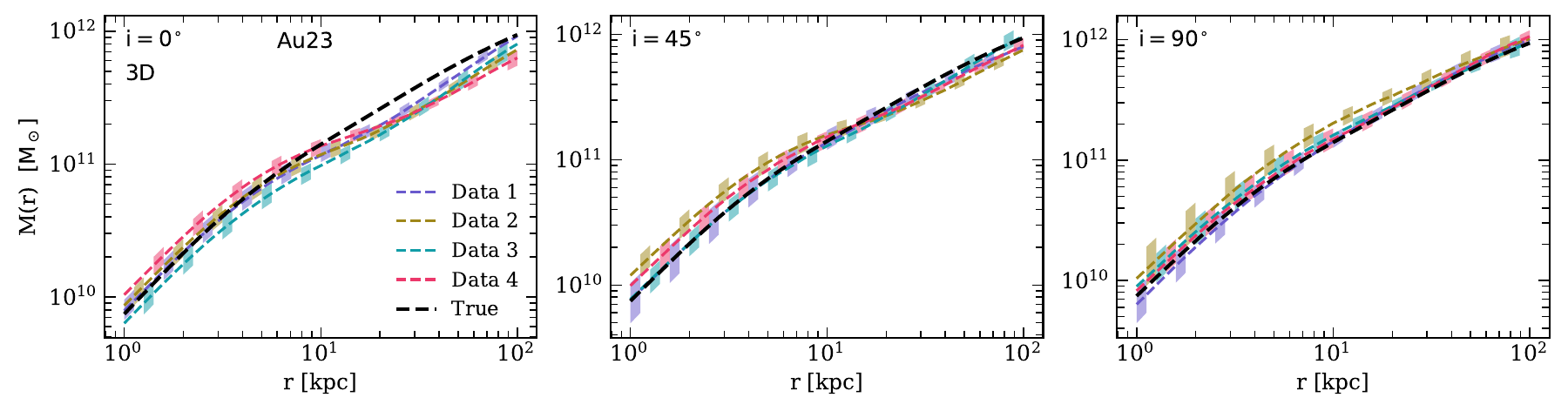}
    \caption{Total enclosed mass of the best-fit models vs true density of the Au23 halo for the (a) ${i=0^{\circ}}$, (b) ${i=45^{\circ}}$, and (c) ${i=90^{\circ}}$ inclinations, where the tracers used to constrain the potential have 3D phase-space information. The different coloured lines correspond to fits to four independent data realisations. The shaded regions are the 1$\sigma$ uncertainties.}
    \label{fig:encmass_total_Au23}
\end{figure*}

The total enclosed mass recovered is compared to the true total mass distribution in Figure \ref{fig:encmass_total_Au23}.

The total mass distribution profiles are well recovered within the confidence intervals for inclinations of $i=45^{\circ}$ and $i=90^{\circ}$. For the face-on case ($i=0^{\circ}$), the recovery remains accurate within $r \approx 10$ kpc, where the baryonic potential dominates, but becomes underestimated at larger radii ($r > 10$ kpc). Nevertheless, the total enclosed mass within 100 kpc is well recovered for all inclination angles. 

\begin{figure*}
    \centering
    \includegraphics[width=1\textwidth]{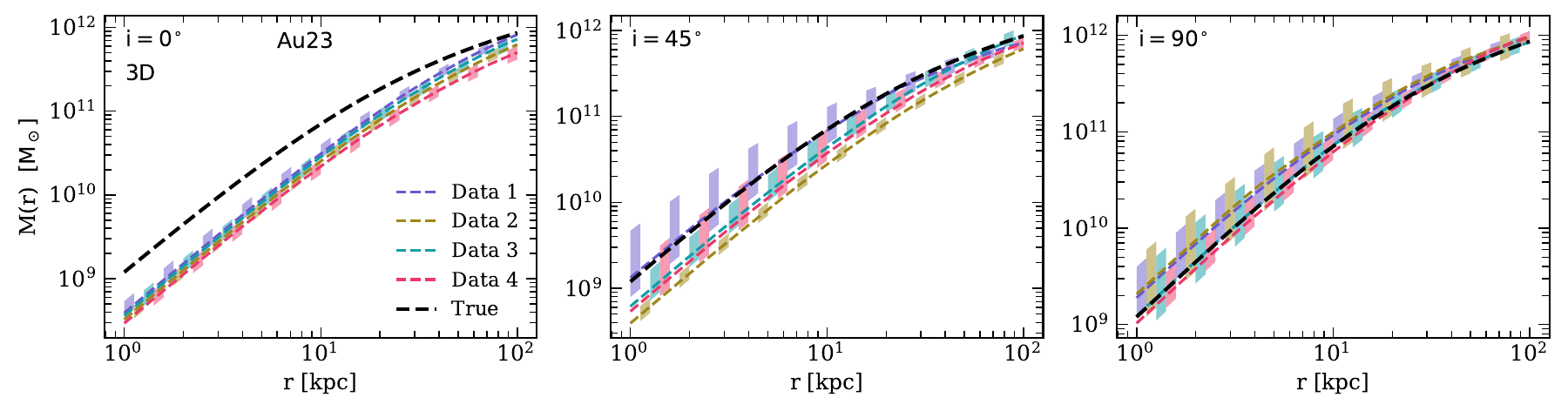}
    \caption{DM enclosed mass of the best-fit model vs true density of the Au23 halo for the (a) ${i=0^{\circ}}$, (b) ${i=45^{\circ}}$, and (c) ${i=90^{\circ}}$ inclinations, where the tracers used to constrain the potential have 3D phase-space information. The different coloured lines correspond to fits to four independent data realisations. The shaded regions are the 1$\sigma$ uncertainties.}
    \label{fig:encmass_dm_Au23}
\end{figure*}

The recovery of the DM mass distribution is shown in Figure \ref{fig:encmass_dm_Au23}. The profiles are well recovered within the confidence intervals for inclinations of $i=45^{\circ}$ and $i=90^{\circ}$, although the associated uncertainties are larger, particularly in the inner regions of the galaxy. For the face-on case ($i=0^{\circ}$), the DM mass distribution is systematically underestimated across all radii, with the bias being most pronounced in the central regions. Nevertheless, the total enclosed DM mass at $r=100$ kpc is recovered within the confidence intervals and with higher accuracy. However, increased uncertainty in the inner DM profile is expected, as the gravitational potential in the central regions reflects the combined influence of the bulge, disc, and DM components, leading to degeneracies in how their respective contributions are inferred. The same trend is observed in our idealised mocks as well.

\par
\begin{figure}
    \centering
    \includegraphics[width=1\linewidth]{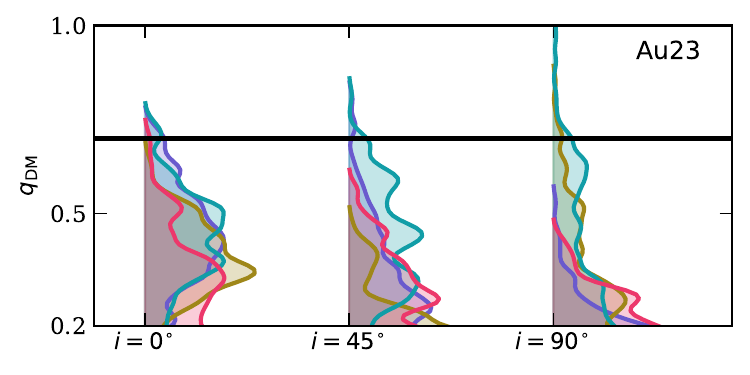}
    \caption{The posterior distribution of the DM flattening parameter $q_\mathrm{DM}$ for all inclinations and data set realisations (coloured lines) for the Au23 halo: ${i=0^{\circ}}$, ${i=45^{\circ}}$, and ${i=90^{\circ}}$, where the tracers used to constrain the potential have 3D phase-space information. The different coloured lines correspond to fits to four independent data realisations. The solid black line is the true value of the $q_\mathrm{DM}$ parameter as recovered in \citealt{Gherghinescu+2024}. }
    \label{fig:q_dm_Au23}
\end{figure}

 The recovery of the flattening of the DM halo is even more challenging in the realistic Auriga 23 case, as it can be seen in Figure \ref{fig:q_dm_Au23}, especially the closer we get to the edge-on inclination $i=90^{\circ}$. We observe a systematic bias towards smaller $q_\mathrm{DM}$ for all datasets and inclinations, as well as a broad $q_{\mathrm{DM}}$ posterior.

\subsection{Kinematics of the tracers}

\begin{figure*}
  \centering
  \includegraphics[width=\linewidth]{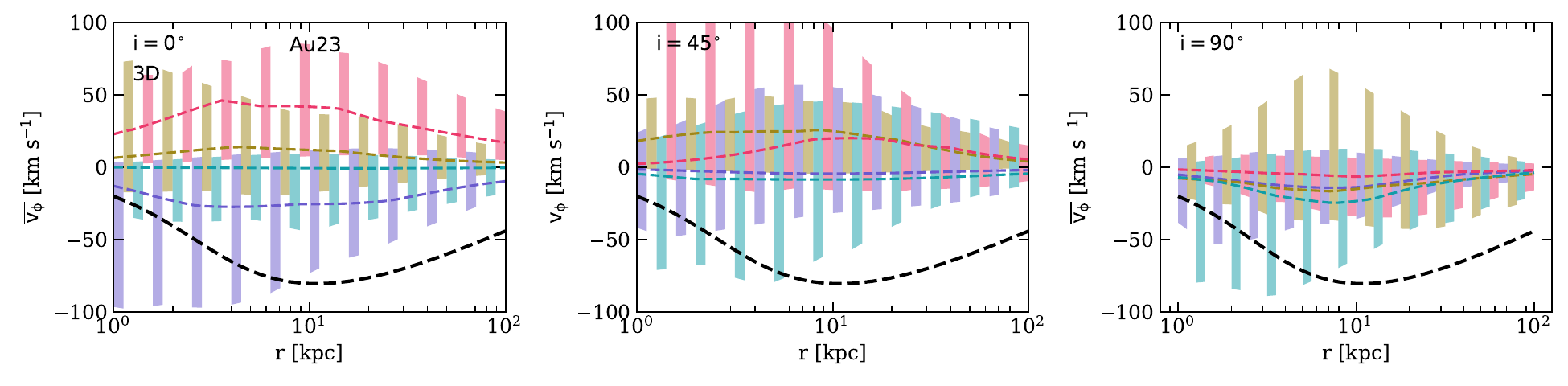}
  \caption{The true vs recovered median ${v_{phi}}$ velocity component (streaming motion) for the Auriga 23 galaxy for (a) ${i=0^{\circ}}$, (b) ${i=45^{\circ}}$, and (c) ${i=90^{\circ}}$, where the tracers used to constrain the potential have 3D phase-space information. The different coloured lines correspond to fits to four independent data realisations. This quantifies the overall degree of rotation about the axisymmetry axis of the stellar halo at different radii. It can be seen that it is best recovered for edge-on orientation and no rotation signal is detected for the face on orientation, as expected.}
  \label{fig:med_vel_vphi_Au23}
\end{figure*}

Figre \ref{fig:med_vel_vphi_Au23} shows the recovery of the streaming rotational motion ($\overline{v_{\phi}}$). This is recovered with larger uncertainty for the $i=0^{\circ}$ and $i=45^{\circ}$ inclinations. In all three inclination cases, the streaming motion is underestimated.

\begin{figure*}
  \centering
    \begin{subfigure}{0.95\linewidth}
    \centering
      \includegraphics[width=\linewidth]{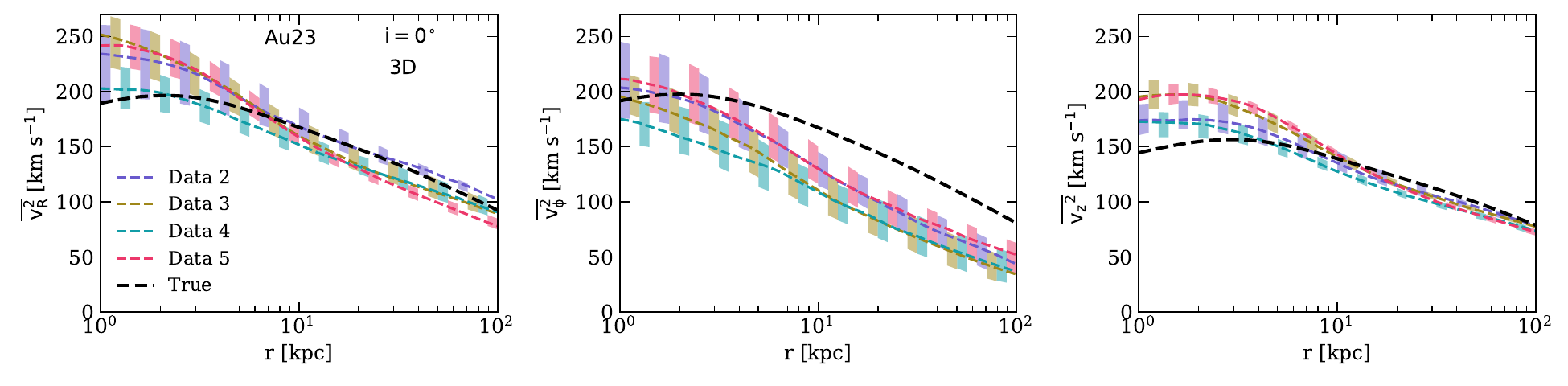}
      \caption{}
      \label{subfig:veldisp_i0_Au23}
    \end{subfigure}\\
    \begin{subfigure}{0.95\linewidth}
        \centering
      \includegraphics[width=\linewidth]{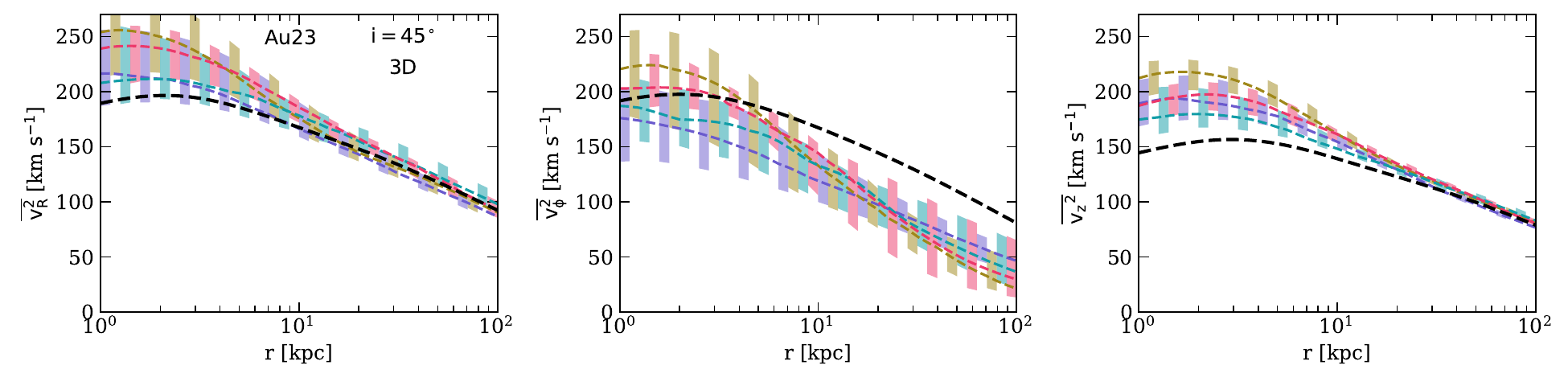}
      \caption{}
      \label{subfig:veldisp_45_Au23}
    \end{subfigure}\\
    \begin{subfigure}{0.95\linewidth}
        \centering
      \includegraphics[width=\linewidth]{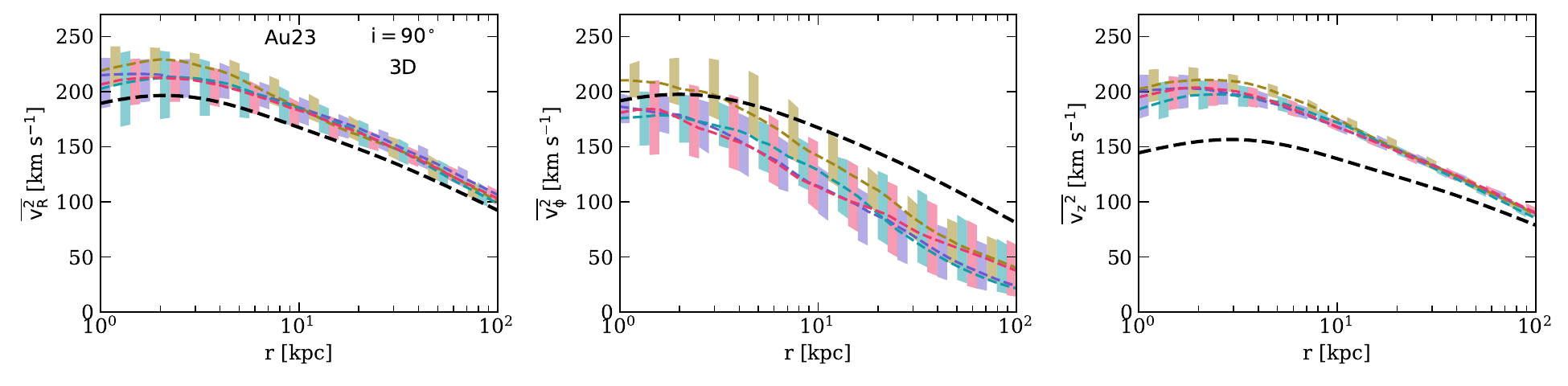}
      \caption{}
      \label{subfig:veldisp_i90_Au23}
    \end{subfigure}    
  \caption{Velocity dispersions in cylindrical coordinates of the recovered models (coloured lines) compared to the true values of the Auriga 23 galaxy (black dotted line) for inclination angles (a) ${i=0^{\circ}}$, (b) ${i=45^{\circ}}$, and (c) ${i=90^{\circ}}$, where the tracers used to constrain the potential have 3D phase-space information. The different coloured lines correspond to fits to four independent data realisations. Shaded regions indicate the 1$\sigma$.}
  \label{fig:veldisp_Au23}
\end{figure*}

Figure \ref{fig:veldisp_Au23} shows the recovery of the velocity dispersions in cylindrical coordinates for the Au23 accreted stellar halo. The same recovery trends and biases are observed as in the mock tests. The vertical dispersion profile $\overline{v_{z}^{2}}$ is best recovered for the face-on inclination ($i=0^{\circ}$), as that is when the $v_{z}$ velocity vector component is aligned with the LOS. The opposite is true for the edge-on inclination ($i=90^{\circ}$). 

\par For the velocity components lying in the axisymmetry plane ($v_{R}$ and $v_{\phi}$), the opposite trend is observed, as expected. The $\overline{v_{R}^{2}}$ and $\overline{v_{\phi}^{2}}$ profiles are more accurately recovered within the confidence intervals for the edge-on inclination ($i=90^{\circ}$), when these components are aligned with the LOS, and show the largest bias and uncertainty in the face-on configuration ($i=0^{\circ}$).

\par Even though similar systematic trends are observed in the recovered kinematic properties of the tracers in both the Auriga 23 halo and idealised mocks, there is a larger degree of uncertainty and bias in the more realistic Auriga 23 halo. A higher level of bias is expected when applying the dynamical model to the cosmological simulation compared to the idealised mock galaxy from Section~\ref{sec:mock_tests}. This is because Auriga 23 represents a more complex and realistic system, featuring a richer dynamical structure, non-equilibrium features, and a wider range of stellar orbits. Moreover, the true mass distributions of both the baryons and the DM often deviate from the idealised analytical profiles assumed in such models, even if these deviations are not dominant at first order.

\section{Schwarzschild modelling}  \label{sec:schwarzschild}

To demonstrate that the inability to constrain the halo flattening $q_\mathrm{DM}$ cannot be explained by some limitations of the DF-based models, we conducted another series of tests, using a completely different modelling approach, namely the Schwarzschild orbit-superposition method \citep{Schwarzschild1979}, as implemented in the \textsc{Forstand} code \citep{Vasiliev&Valluri2020}. Instead of fitting a parameterised DF to $\mathcal O(10^3)$ discrete kinematic tracers, we constructed 3d kinematic datacubes with LOS velocity distributions $f(v_Z)$ on a grid in the $X,Y$ image plane. This polar grid had 30 radial bins equally spaced in the enclosed mass up to a radius 100~kpc containing over 99\% of the stellar halo tracers, and 4 angular bins covering one quadrant between major and minor axes. The $v_\mathrm{LOS}$ distributions are represented by Gauss--Hermite series \citep{vanderMarel&Franx1993,Gerhard1993} up to 8th order, which is larger than used in most dynamical studies (see \citealt{Liepold+2020} for a discussion), and the surface density of tracers is constrained by the overall normalisation of the LOS velocity distribution in each bin. In total, we have $30\times4\times9$ constraints in the Schwarzschild model, and their uncertainties are obtained by bootstrapping over independent realizations of $10^6$ samples drawn from the true DF of the model, reaching $\sim 1\%$ for the surface density, $\sim 1$~km\,s$^{-1}$ for $v$ and $\sigma$, and $\sim 0.005$ for the higher-order Gauss--Hermite moments. Already the total number of constraints is larger than the $10^3$ stars used in the discrete-kinematics DF models; although it is not straightforward to compare the constraining power of discrete tracers vs.\ binned velocity distributions, it is clear that the latter ones are much tighter.

For simplicity, we fix all parameters of the potential except the flattening $q_{\mathrm{DM}}$ and the overall mass normalisation $\Upsilon$; if anything, this would restrict the range of models consistent with the observed tracer kinematics more strongly than in our DF models. Despite this, we find that the orbit-superposition models can fit the data nearly perfectly in a large range of $q_{\mathrm{DM}}$ values. For the face-on case ($i=0^\circ$), any value of $q_{\mathrm{DM}}$ between 0.2 and 1 is consistent with the data; for the intermediate inclination ($i=45^\circ$), the allowed range of $q_{\mathrm{DM}}$ is between 0.4 and 1; and in the edge-on case ($i=90^\circ$), we could additionally exclude $q_{\mathrm{DM}}>0.8$ (we recall that the true value of $q_{\mathrm{DM}}=0.7$). This confirms the trends we have seen with the DF-based models: the constraints on the potential shape are fairly weak in the edge-on case, and practically disappear as the inclination decreases to face-on case, demonstrating that this is a genuine degeneracy in dynamical models, independent of the specific choice of modelling method.

\section{Discussion and conclusions} 
\label{sec:discussion_conclusions}
In this work, we have built upon the action-based dynamical model in \cite{Gherghinescu+2024} to develop a method for constraining the total mass and DM distribution external galaxies with reduced phase space distribution availability, as well as constrain the phase-space distribution of the stellar halo component. In \cite{Gherghinescu+2024} we have shown that the equilibrium action-based dynamical models provide a very good fit to the galactic visible and DM mass distribution, given full phase-space information of the tracer population. In this paper, we investigated the effects of missing phase-space information.

\par Both our idealised mock and Au23 simulation tests show that additional biases appear in the recovery of the total and DM mass distributions of the galaxies when the data is degraded. We have seen that while this effect comes from the missing phase-space information of the tracers, inclination also plays a role. 

\par For the recovery of the kinematics of the tracer population, both the idealised mocks and simulations tests have shown that the velocity components that align with the LOS are better recovered, which is reflected in our tests. The ${v_{z}}$ velocity component in the intrinsic coordinate system of the galaxy is best recovered when the galaxy is viewed face-on ($i=0^{\circ}$) and is most biased when it is viewed edge-on $i=90^{\circ}$. The in-plane (of the disc) velocity components ${v_{R}}$ and ${v_{\phi}}$ show the opposite trend to ${v_{z}}$, which is expected.

\subsection{Mock tests on 4D and 5D phase-spae information} \label{sub:4d5d6d_mock}
The goal of this paper is to assess how much we can learn about the total mass distribution and the phase-space structure of tracers using only their on-sky positions and LOS velocities. However, it also insighful to systematically track how the progressive loss of phase-space information impacts the quality of dynamical fits. 

We ran numerous experiments with different amounts of data available for the fit, using the same mock galaxy as in Section \ref{sec:mock_tests}.  We started from the case of full 6D phase-space information, then removing the distance ($Z$ coordinate, i.e., retaining 5D information), and then removing either one ($v_Z$, 4D) or two ($v_{X,Y}$, 3D) velocity components. 

\par
\begin{figure*}
  \centering
    \begin{subfigure}{0.95\linewidth}
    \centering
      \includegraphics[width=\linewidth]{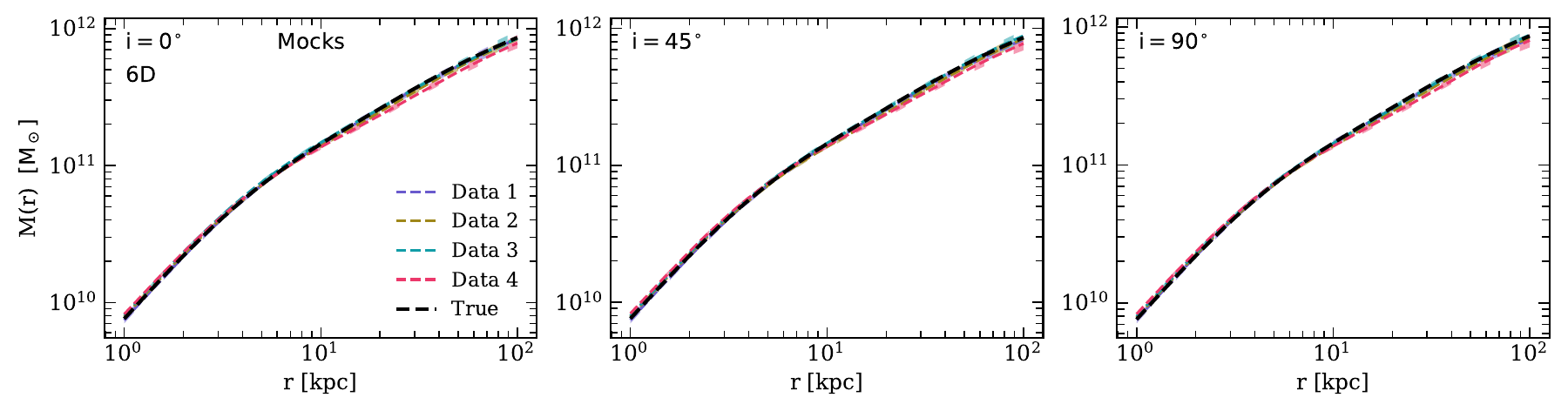}
      \caption{}
      \label{subfig:enclosed_mass_total_6D}
    \end{subfigure}\\
    \begin{subfigure}{0.95\linewidth}
        \centering
      \includegraphics[width=\linewidth]{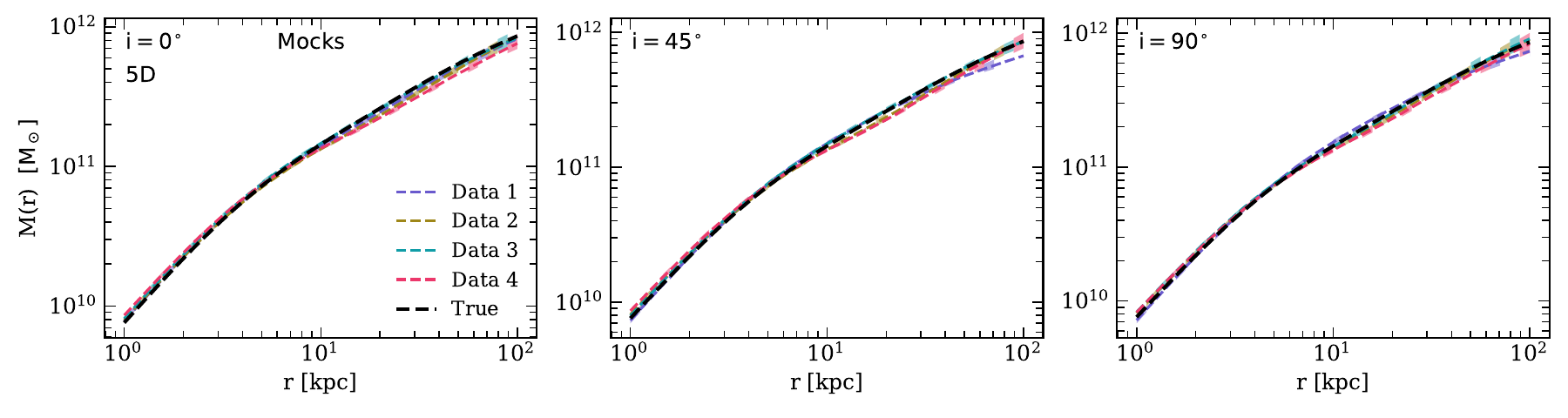}
      \caption{}
      \label{subfig:enclosed_mass_total_5D}
    \end{subfigure}\\
    \begin{subfigure}{0.95\linewidth}
        \centering
      \includegraphics[width=\linewidth]{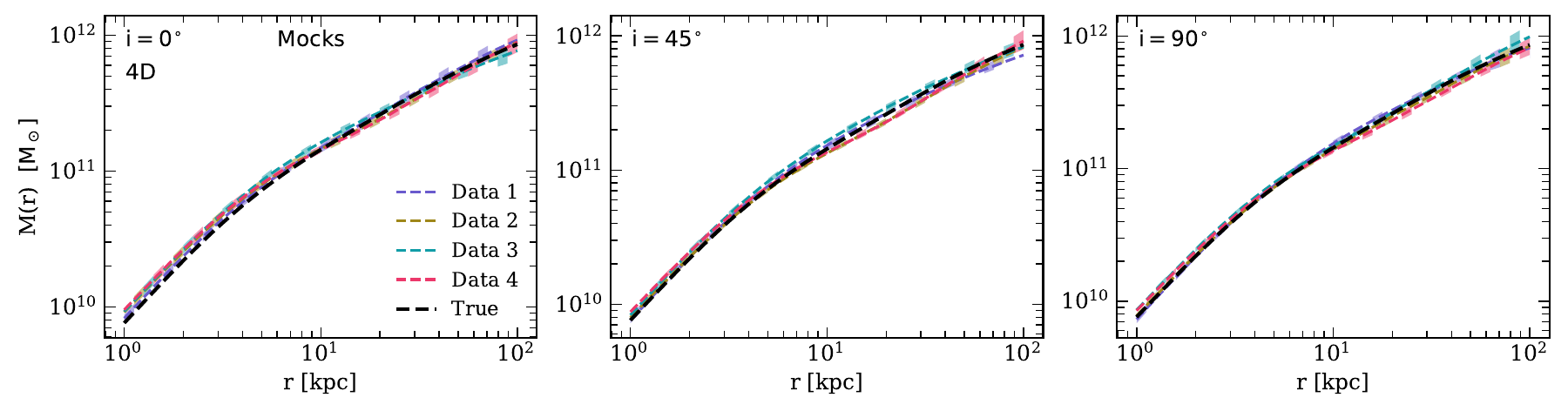}
      \caption{}
      \label{subfig:enclosed_mass_total_4D}
    \end{subfigure}    
  \caption{Total enclosed mass profiles of the best-fit models (coloured lines) compared to the true mass distribution of the mock galaxy (black dotted line) for inclination angles, where the tracers used to constrain the potential have (a) 6D (full), (b) 5D, and (c) 4D phase-space information. The different coloured lines correspond to fits to the same four independent data realisations as in Section \ref{sec:mock_tests}.  Shaded regions indicate the 1$\sigma$ uncertainties.}
  \label{fig:enclosed_mass_total_4D_5D_6D}
\end{figure*}

\begin{figure*}
  \centering
    \begin{subfigure}{0.95\linewidth}
    \centering
      \includegraphics[width=\linewidth]{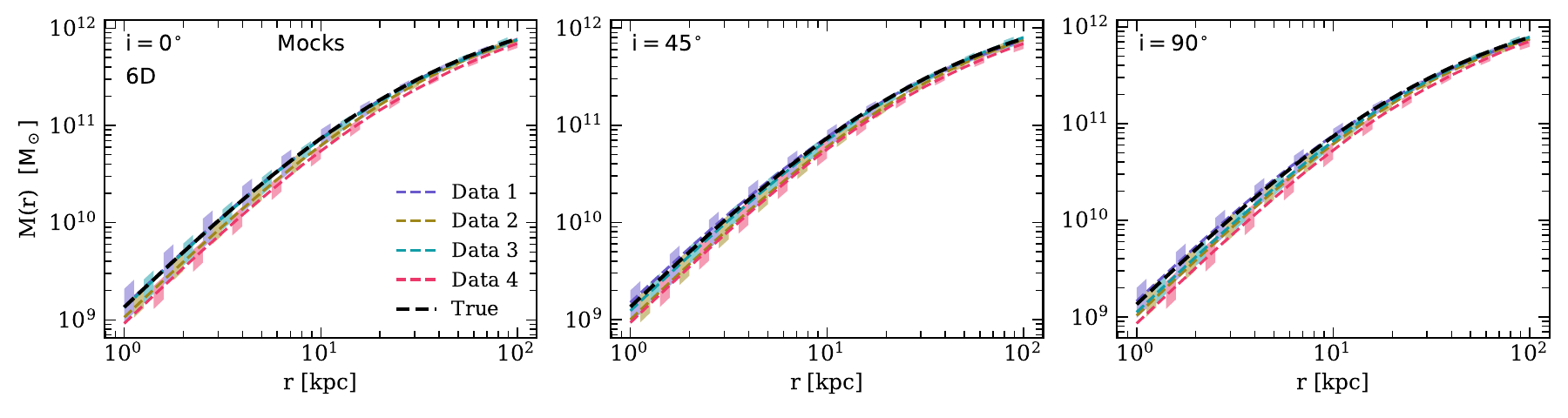}
      \caption{}
      \label{subfig:enclosed_mass_dm_6D}
    \end{subfigure}\\
    \begin{subfigure}{0.95\linewidth}
        \centering
      \includegraphics[width=\linewidth]{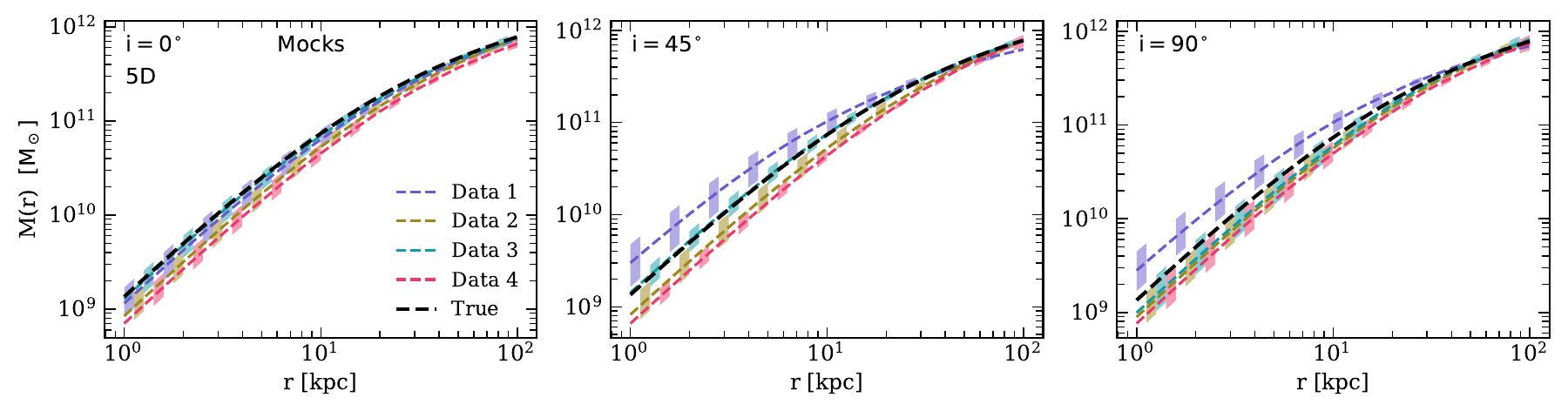}
      \caption{}
      \label{subfig:enclosed_mass_dm_5D}
    \end{subfigure}\\
    \begin{subfigure}{0.95\linewidth}
        \centering
      \includegraphics[width=\linewidth]{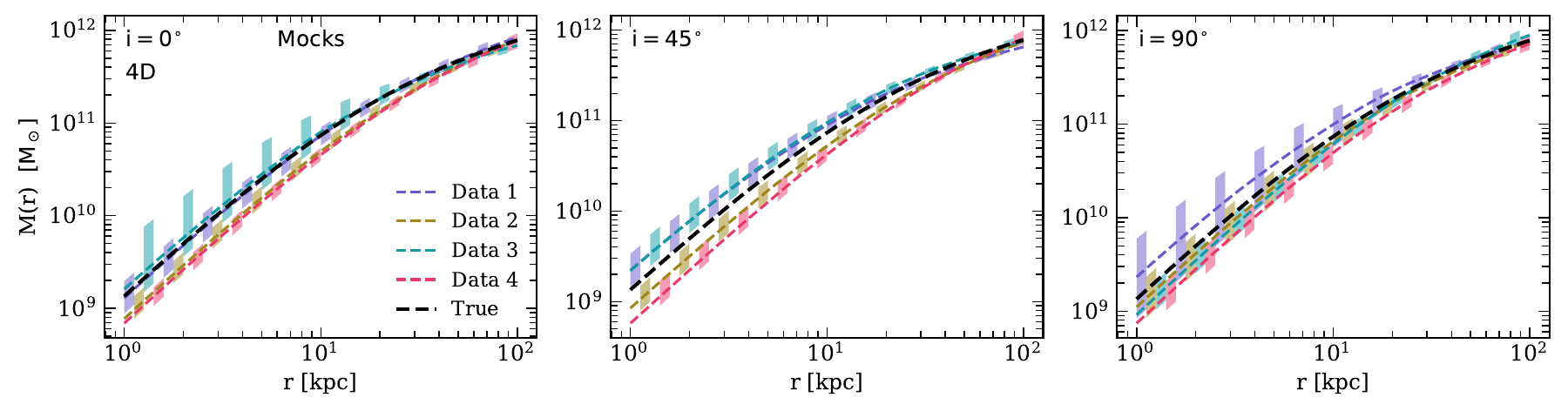}
      \caption{}
      \label{subfig:enclosed_mass_dm_4D}
    \end{subfigure}    
  \caption{DM enclosed mass profiles of the best-fit models (coloured lines) compared to the true mass distribution of the mock galaxy (black dotted line) for inclination angles, where the tracers used to constrain the potential have (a) 6D (full), (b) 5D, and (c) 4D phase-space information. The different coloured lines correspond to fits to the same four independent data realisations as in Section \ref{sec:mock_tests}. Shaded regions indicate the 1$\sigma$ uncertainties.}
  \label{fig:enclosed_mass_dm_4D_5D_6D}
\end{figure*}

\begin{figure*}
    \includegraphics[width=1\linewidth]{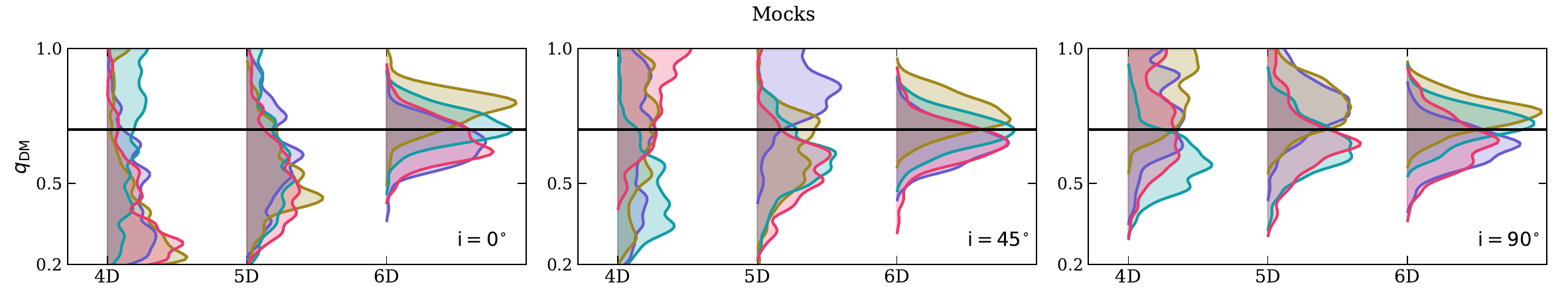}
  \caption{The posterior distribution of the DM flattening parameter $q_\mathrm{DM}$ for all inclinations ($i=0^{\circ}$, $i=45^{\circ}$, and $i=90^{\circ}$, left to right) and data set realisations (coloured lines), where the tracers used to constrain the potential have 6D (full), 5D, and 4D phase space information. The different coloured lines correspond to fits to the same four independent data realisations as in Section \ref{sec:mock_tests}. The solid black line is the true value of the parameter used to generate the mock galaxies (i.e., $q_\mathrm{DM,\,true}=0.7$).}
  \label{fig:q_dm_4D_5D_6D}
\end{figure*}

 Across all 6D, 5D, and 4D configurations, the dynamical fits successfully reproduce the total mass profiles (Figure \ref{fig:enclosed_mass_total_4D_5D_6D}). The DM distribution is also well recovered within the confidence intervals (Figure \ref{fig:enclosed_mass_dm_4D_5D_6D}), though the fits become more uncertain in the inner regions as the available phase-space information is reduced. Figure \ref{fig:q_dm_4D_5D_6D} show that as progressively less phase-space information is available for the potential tracers, the posterior distribution of the recovered DM flattening, $q_{\mathrm{DM}}$, becomes increasingly broad and flat. This indicates that the model becomes effectively insensitive to the true flattening of the DM halo.

\subsection{The limits of the St\"ackel approximation in the stellar halo} \label{sub:stackel_fudge}

The St\"ackel fudge is an \textit{approximate} method for computing orbital actions in an axisymmetric potential. It assumes that the local gravitational potential can be represented as a St\"ackel potential, for which actions can be calculated analytically. By making this approximation, one can estimate the actions of stars efficiently.

\par However, it is possible that the St\"ackel approximation used for action computation is not accurate enough, especially for the more extreme orbits present. As discussed in \citet{Sanders&Binney2016}, this method can fail catastrophically for orbits near or within resonant regions, producing errors of order unity. Furthermore, the St\"ackel approximation can be less accurate for stars on highly eccentric, inclined, or energetic orbits, such as those commonly found in stellar halos. In these cases, the local potential may deviate significantly from a true St\"ackel form along it's orbit. This could explain the systematic accumulation of $q_\mathrm{DM}$ parameter at low values observed in the Auriga 23 fits.

\par Although on average the St\"ackel approximation delivers unbiased action estimates (section~3.3 in \citealt{Vasiliev2019_Agama}), this does not necessarily mean that the values of the likelihood will be unbiased. Indeed, since the DF is proportional to actions raised to some negative power, a symmetrically distributed error in actions will after averaging bias the values of the DF upward. Since the St\"ackel approximation is essentially exact for spherical potentials and its accuracy decreases as one moves towards smaller $q_\mathrm{DM}$, this may explain the observed bias in the fits. There exist other, more accurate methods for action computation, e.g., the O2GF method of \citet{Sanders&Binney2014}; however, in practice we found the existing implementation not to be 100\% reliable, and even a small fraction of incorrectly computed actions can thwart the overall accuracy of likelihood computation. Besides, this approach is considerably more expensive and impractical for large-scale MCMC fits. 

To assess how the Stäckel fudge affects the recovery of $q_{\mathrm{DM}}$, we repeated the DF-fitting experiments using idealised mocks built in a Stäckel potential—the Perfect Ellipsoid \citep{deZeeuw1985}—for which the action computation is exact (see Appendix \ref{appendix:idealised_mocks_perfect_ellipsoid}). We find that in these runs, the posterior distribution of axis ratio $q_\mathrm{DM}$ is still fairly broad (Figure \ref{fig:q_perfect_ellipsoid_3D_4D_5D_6D}), and while it may be shifted to lower or higher values than the true one for different realizations of the mock data, overall it seems to avoid being \textit{systematically biased}. Thus, inaccuracies in the action computation inherent to the St\"ackel fudge cannot account for the broad $q_{\mathrm{DM}}$ posterior or the inability to recover the DM halo flattening.

\subsection{Flattening of the DM halo}
The results this work show that the model’s sensitivity to the true DM halo flattening, $q_\mathrm{DM}$, decreases as phase-space information is reduced, leading to broader and less constrained posterior distribution.
We conclude that recovering the DM halo flattening $q_{\mathrm{DM}}$ is a challenging task, if at all possible, when limited phase space information of the tracers is available.

We recall that in previous studies using the DF fitting approach to infer the gravitational potential, it was either limited to spherical systems \citep{Pascale+2018,Pascale+2019,Pascale+2024,DellaCroce+2024,ArroyoPolonio+2025} or cases with 6D (\citealt{Posti+2019}\footnote{Note that their study used the axisymmetric St\"ackel approximation outside the range of applicability of its current implementation ($q_\mathrm{DM}\le 1$).}, \citealt{Vasiliev2019_clusters}, \citealt{CorreaMagnus+2022}, \citealt{Binney+2023}) or 5D \citep{Hattori+2021} phase-space information, for which the inaccuracies introduced by the St\"ackel approximation are less noticeable.

As our experiments with the Schwarzschild method demonstrate, the impossibility of tightly constraining the halo flattening must be a more general property of dynamical models. This conclusion is mostly supported by the prior literature.
For instance, \citet{Hagen+2019} tested their axisymmetric Schwarzschild modelling machinery in a similar context to ours (stellar tracers in a flattened DM halo potential) and found that while they could successfully recover the mass profile, the halo flattening was poorly constrained. Most applications of Schwarzschild models stipulate that the kinematic tracers (stars) also create the gravitational potential, either entirely or partially (there may be other components in the potential, such as central black hole or a DM halo, usually spherical). In the gravitationally self-consistent case, unlike our experiments, the flattening of the potential is directly inferred from the observed shape of the stellar distribution (if the inclination can be well constrained). \citet{vandenBosch+2009} fitted self-consistent triaxial and (nearly) axisymmetric Schwarzschild models to mock data, and found that the shape was only weakly constrained in the axisymmetric case, but much more strongly in the triaxial case, in which the structure of the velocity field is more complicated (e.g., with kinematic twists). 
On the other hand, \citet{deNicola+2022} and \citet{Neureiter+2023} find that their implementation of the Schwarzschild method, when applied to mock data from a simulation in which both stars and DM halo have triaxial shapes, can recover the total mass profile to within 10\% and the axis ratios of both stars and DM within $\lesssim 0.1$. \citet{Lipka+2024}, using the same code but with an axisymmetric halo, also found that the models correctly recover the true axis ratio (0.8 in their case) when projected along the intermediate axis (i.e. edge-on), strongly excluding a spherical halo. We recall that in our tests, we could also place an upper limit on the axis ratio ($q_\mathrm{DM}\lesssim 0.9$) only in the edge-on orientation. It remains to be explored why their tests suggest a better precision in constraining the halo shape than ours, given that their stellar tracers represent elliptical galaxies and are more centrally concentrated relative to the DM halo than in our experiments. 

Our findings suggest that when only 3D phase-space information of tracers is available, one should not expect to obtain strong and reliable constraints on the flattening of the DM halo from equilibrium models. In our experiments, the accuracy of recovering the mass profile is mostly unaffected by the flattening of the potential, so the simplest approach would be to continue using spherical haloes, as done in most previous studies. 
On the other hand, the morphology of stellar tidal streams can provide meaningful constraints on the radial profile and the shape of the DM halo even in absence of kinematic measurements \citep{Nibauer+2023,Walder+2024}, but is insensitive to the absolute mass normalisation. Thus combining both streams and equilibrium modelling may offer a path for breaking the degeneracies in halo properties.

\subsection{Concluding remarks}
We presented an axisymmetric, action-based dynamical modelling method to constrain the total and DM mass profiles in external galaxies using resolved stellar halo tracers (with 3D phase-space information, i.e., on-sky positions and LOS velocities). Importantly, \textrm{both} the tracer DF and the gravitational potential in our models are fully axisymmetric and not related to each other (i.e., tracers do not contribute to the potential).

\par We provide very extensive tests to illustrate biases introduced by missing phase space information of the stellar tracers, inclination effects, as well as those introduced by our methodology (marginalisation integral computation, the St\"ackel fudge action computation method, the equilibrium assumption, etc.). Despite these, our projected DF method does well in recovering the mass distribution of the galaxies, both in idealised equilibrium mocks and in realistic cosmological hydrodynamical simulations. However, we found that the flattening of the DM halo cannot be meaningfully constrained when only 3D or 4D phase-space information is available. This is likely a fundamental degeneracy, rather than a limitation of the DF-fitting method, as confirmed by additional tests using a different modelling approach (Schwarzschild orbit-superposition method).

\par We have improved the MC marginalisation integral computation (see Equation \ref{eq:marginalisation}) presented in \citealt{Read+2021} in several ways: more efficient deprojection, optimised importance sampling of velocity components, and the use of low-discrepancy quasi-random numbers (see Section~\ref{sec:marginalisation} for details).

Our work demonstrates that action-based DF models are powerful even in regimes with few tracers or incomplete phase-space information. They avoid binning, utilize the full velocity distribution rather than just its first two moments, and can be used with anisotropic and non-spherical (axisymmetric) stellar systems. Although alternative methods based on Jeans equations, such as \texttt{CJAM} \citep{Watkins+2013} and \texttt{GravSphere2} \citep{BanaresHernandez+2025}, can deal with unbinned data, non-spherical systems, and higher-order velocity moments, none of the current implementations satisfies all these requirements, unlike our DF-fitting approach.

The method presented in this work is particularly well suited for nearby systems such as the Andromeda galaxy (M31), whose favourable inclination on the sky and availability of individual stellar tracers (e.g. planetary nebulae and resolved halo stars) provide ideal conditions for applying our dynamical modelling framework to uncover its mass distribution and stellar halo phase space distribution. We leave this application for future work.

\section*{Acknowledgements}
EV acknowledges support from an STFC Ernest Rutherford fellowship (ST/X004066/1). PD is supported by a UKRI Future Leaders Fellowship (grant reference MR/S032223/1).

\section*{Data Availability}

The Auriga simulations and Agama software are publicly available. A baseline example code for DF-based dynamical modelling is included in the Agama repository.



\bibliographystyle{mnras}
\bibliography{bibliography} 




\appendix

\section{Perfect ellipsoid idealised mocks test} \label{appendix:idealised_mocks_perfect_ellipsoid}
In this test, we repeat the setup described in Section \ref{sec:mock_tests}, but replace the potential with a simpler, single-component perfect ellipsoid, which is of St\"ackel form. In such potentials, the action computation is exact (performed by 1d numerical quadrature with sufficient precision). The perfect ellipsoid density profile is
\begin{equation}
    \rho(R,z) = \frac{M_\mathrm{DM}}{\pi^{2}\, q_\mathrm{DM}\, R_\mathrm{DM}^{3}} \bigg( 1+ \frac{R^{2} + (z/q_\mathrm{DM})^{2}}{R_\mathrm{DM}^{2}} \bigg)^{-2} ,
    \label{eq:perfect_ellipsoid}
\end{equation}
while the DF is of the same double-power-law family as in Eq.~\ref{eq:double_power_law_df}. The details of the model can be found in Table~\ref{tab:mock_param_perfect ellipsoid}.

\begin{figure*}
  \centering
    \begin{subfigure}{0.95\linewidth}
    \centering
      \includegraphics[width=\linewidth]{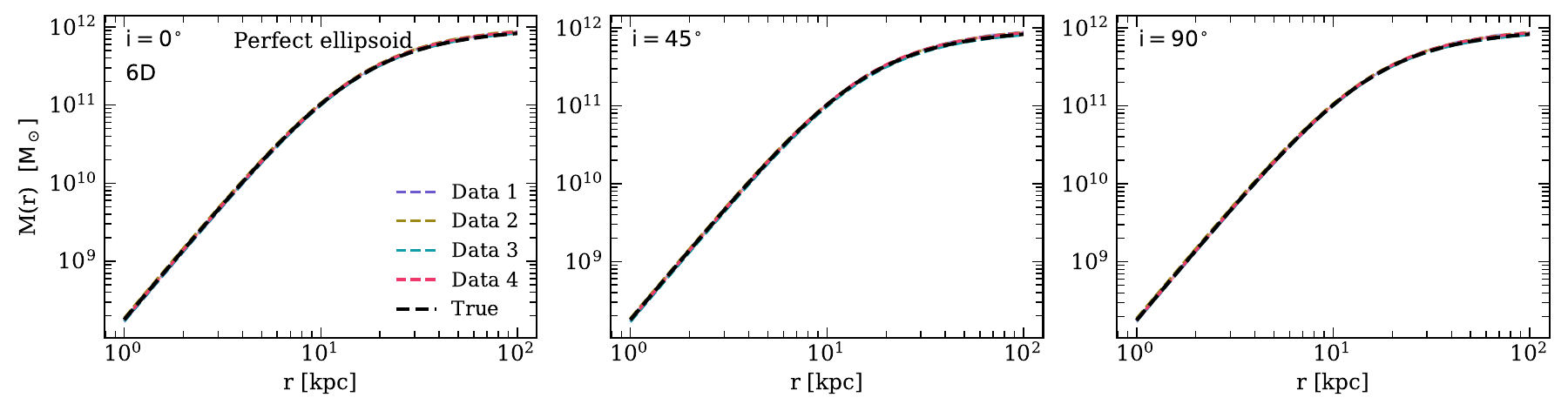}
      \caption{}
      \label{subfig:enclosed_mass_perfect_ellipsoid_6D}
    \end{subfigure}\\
    \begin{subfigure}{0.95\linewidth}
        \centering
      \includegraphics[width=\linewidth]{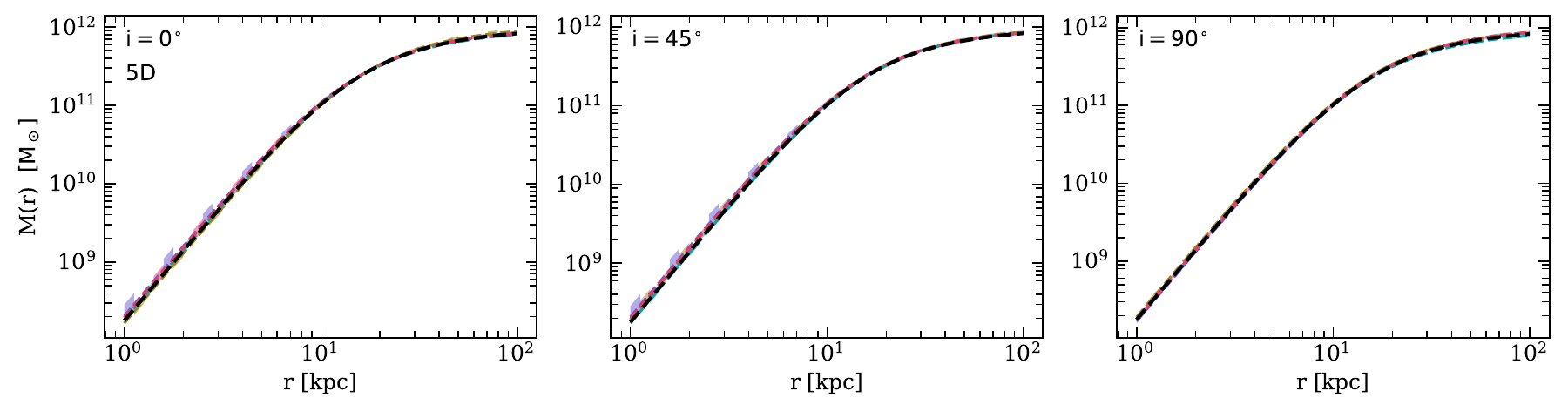}
      \caption{}
      \label{subfig:enclosed_mass_perfect_ellipsoid_5D}
    \end{subfigure}\\
    \begin{subfigure}{0.95\linewidth}
        \centering
      \includegraphics[width=\linewidth]{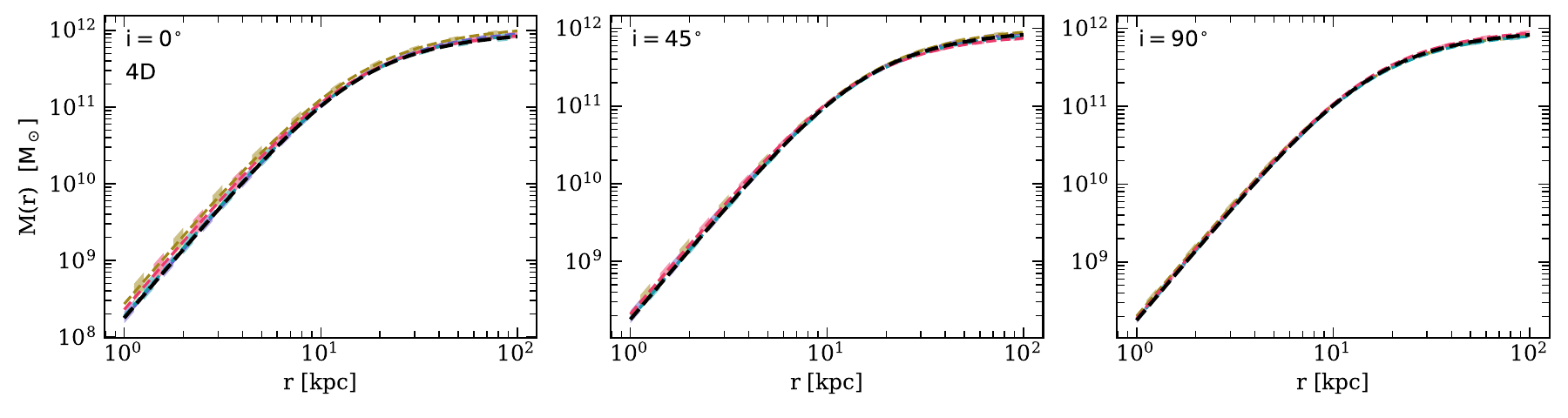}
      \caption{}
      \label{subfig:enclosed_mass_perfect_ellipsoid_4D}
    \end{subfigure}
        \begin{subfigure}{0.95\linewidth}
        \centering
      \includegraphics[width=\linewidth]{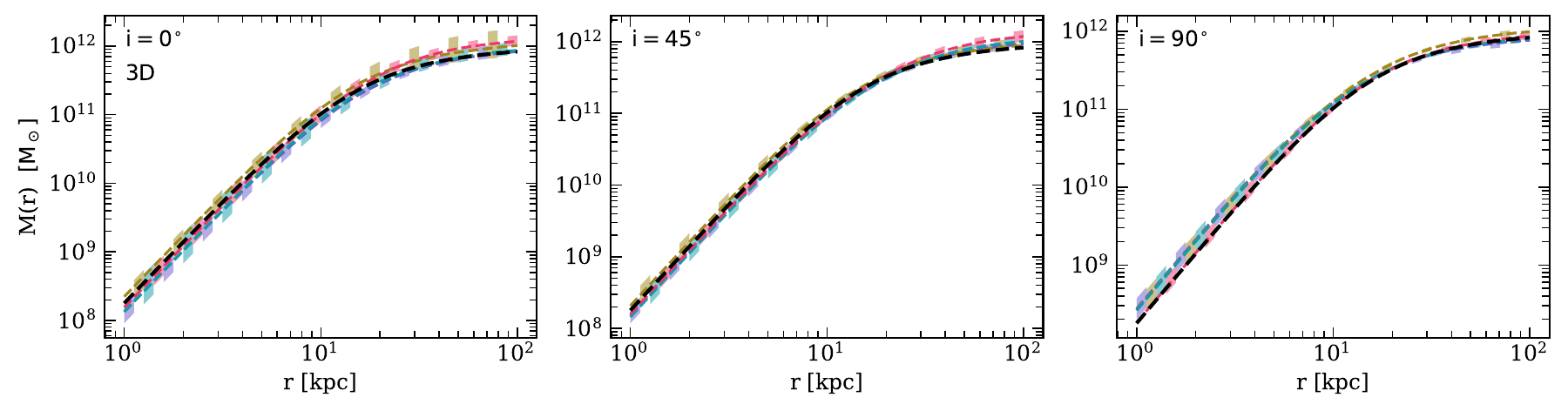}
      \caption{}
      \label{subfig:enclosed_mass_perfect_ellipsoid_3D}
    \end{subfigure}  
  \caption{Total enclosed mass profiles of the best-fit models (coloured lines) compared to the true mass distribution of the perfect ellipsoid mock test galaxy (black dotted line) all for inclination angles, where the tracers used to constrain the potential have (a) 6D (full), (b) 5D, and (c) 4D, and (d) 3D phase-space information. The different coloured lines correspond to fits to the same four independent data realisations as in Section \ref{sec:mock_tests}.  Shaded regions indicate the 1$\sigma$ uncertainties.}
  \label{fig:enclosed_mass_perfect_ellipsoid_3D_4D_5D_6D}
\end{figure*}

\begin{figure}
  \centering
    \begin{subfigure}{0.95\linewidth}
    \centering
      \includegraphics[width=0.9\linewidth]{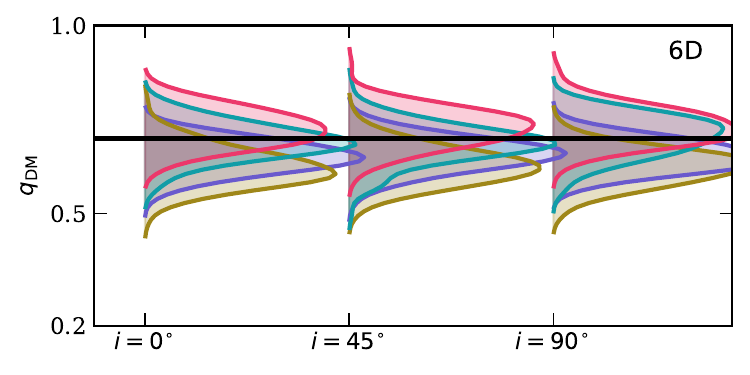}
      \caption{}
      \label{subfig:q_perfect_ellipsoid_6D}
    \end{subfigure}\\
    \begin{subfigure}{0.95\linewidth}
        \centering
      \includegraphics[width=0.9\linewidth]{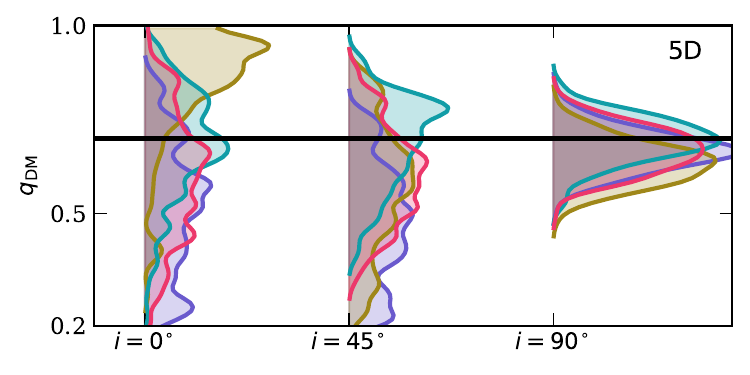}
      \caption{}
      \label{subfig:q_perfect_ellipsoid_5D}
    \end{subfigure}\\
    \begin{subfigure}{0.95\linewidth}
        \centering
      \includegraphics[width=0.9\linewidth]{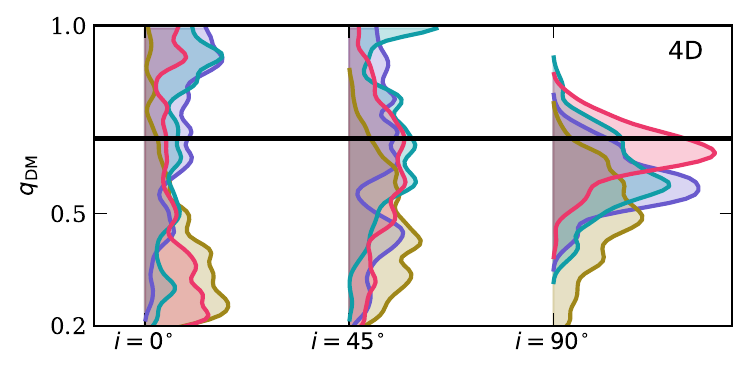}
      \caption{}
      \label{subfig:q_perfect_ellipsoid_4D}
    \end{subfigure}
        \begin{subfigure}{0.95\linewidth}
        \centering
      \includegraphics[width=0.9\linewidth]{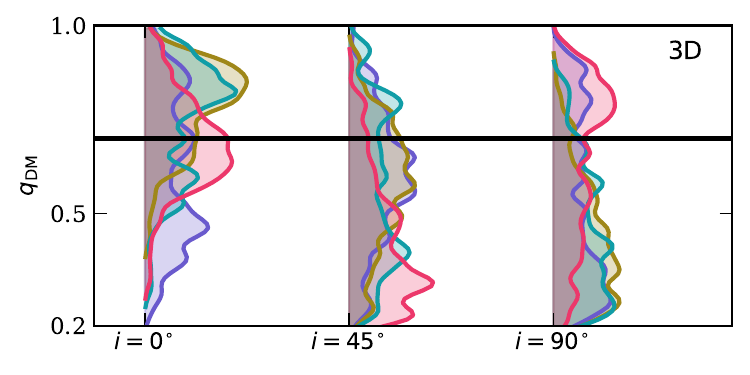}
      \caption{}
      \label{subfig:q_perfect_ellipsoid_3D}
    \end{subfigure}  
  \caption{Posterior distributions of the DM flattening parameter $q_{\rm DM}$ for the perfect ellipsoid mock test, shown for all inclinations and data set realizations (colored lines). Tracers used to constrain the potential have (a) full 6D, (b) 5D, (c) 4D, and (d) 3D phase-space information. Different colors indicate four independent data realizations, and the solid black line shows the true value used to generate the mock galaxy ($q_{\rm DM,true} = 0.7$).}
  \label{fig:q_perfect_ellipsoid_3D_4D_5D_6D}
\end{figure}

As in Section~\ref{sec:mock_tests}, we create four independent realisations of mock datasets, observed at three inclinations, and having between 3 and 6 phase-space coordinates available. Figure~\ref{fig:enclosed_mass_perfect_ellipsoid_3D_4D_5D_6D} demonstrates that the enclosed mass profile is well recovered in all cases (with progressively larger uncertainties as the amount of phase-space information is reduced). The inclination is also recovered to within $\sim$5--10$^\circ$ for 6D and 5D cases, and within 20--30$^\circ$ for 4D and 3D cases.
Figure~\ref{fig:q_perfect_ellipsoid_3D_4D_5D_6D}, on the other hand, shows that the axis ratio $q_\mathrm{DM}$ of the potential is strongly constrained only in 6D or in the edge-on 5D cases, and is somewhat constrained in the edge-on 4D case. In other situations, the posterior distributions of $q_\mathrm{DM}$ are fairly broad, and on average do not show significant systematic biases towards higher or lower values (although they may be quite asymmetric for any particular realisation). This demonstrates that the degeneracy in recovering $q_\mathrm{DM}$ is an inherent feature of dynamical models with incomplete phase-space information, and is largely insensitive to whether the action computation is exact or approximate.

\begin{table}
  \caption{\label{tab:mock_param_perfect ellipsoid} The parameters values of the mock galaxy for the perfect ellipsoid test.}
  \centering 
  \begin{threeparttable}
    \begin{tabular*}{0.8\linewidth}{@{\extracolsep{\fill}}ccc}
\midrule
    Component  & Parameter & Value mock \\
     \midrule
Perfect ellipsoid potential  &   
$\begin{aligned}
    M_\mathrm{DM}\\
    R_\mathrm{DM} \\
    q_\mathrm{DM}
\end{aligned}
$        &  
$\begin{aligned}
    & 1\times 10^{12}\: M_{\odot} \\
    & 15 \textrm{ kpc} \\
    & 0.7
\end{aligned}$ 

\\
    \cmidrule(l  r ){1-3}
     Stellar halo DF & $ \begin{aligned}
    \alpha \\
    \beta \\
    \eta \\
    J_{0} \\
    h_{r} \\
    h_{z} \\
    g_{r} \\
    g_{z} \\
    \chi \\
    J_{\phi,0}
    \end{aligned} $ &
    $\begin{aligned} 
    & 2.11\\
    & 6 \\
    & 1 \\
    & 6000 \: \mathrm{kpc\,km\,s^{-1}} \\
    & 1 \\
    & 1.3 \\
    & 1 \\
    & 1.3 \\
    & 0.5 \\
    & 1000 \: \mathrm{kpc\,km\,s^{-1}} \\
    
    \end{aligned}$  

\\ 
    \midrule
    \end{tabular*}
\end{threeparttable}

\end{table}

\bsp	
\label{lastpage}
\end{document}